\documentclass[conference]{IEEEtran}
\usepackage{amsmath,amssymb,amsfonts}
\usepackage{algorithmic}
\usepackage{graphicx}
\usepackage{textcomp}
\usepackage{xcolor}

\usepackage{amsmath,amssymb,amsfonts}
\usepackage{algorithmic}
\usepackage{graphicx}
\usepackage{textcomp}
\usepackage{xcolor}

\usepackage{fancyhdr}
\usepackage[normalem]{ulem}
\usepackage[bookmarks=true,breaklinks=true,colorlinks,linkcolor=black,citecolor=blue,urlcolor=black]{hyperref}
\usepackage{tabularx}
\usepackage{booktabs}
\usepackage{multirow}
\usepackage{siunitx}
\usepackage{verbatim}
\usepackage{xspace}
\usepackage{enumitem}
\usepackage[linesnumbered,ruled,vlined]{algorithm2e}
\usepackage{dblfloatfix}
\usepackage{multicol}
\usepackage{bbding}
\usepackage[switch]{lineno}

\def\BibTeX{{\rm B\kern-.05em{\sc i\kern-.025em b}\kern-.08em
    T\kern-.1667em\lower.7ex\hbox{E}\kern-.125emX}}

\graphicspath{{figures/plots_arXiv_v1/}}

\newcommand{\MSstat}{$MS_{stat}$\xspace}
\newcommand{\MSdyn}{$MS_{dyn}$\xspace}

\newcommand{\red}[1]{\textcolor{black}{#1}}
\newcommand{\blue}[1]{\textcolor{black}{#1}}
\newcommand{\white}[1]{\textcolor{white}{#1}}

\newcommand{\SCHED}{HetSched\xspace}
\newcommand{\META}{\texttt{Meta-Sched}\xspace}
\newcommand{\TASK}{\texttt{Task-Sched}\xspace}
\newcommand{\DET}{DET\xspace}
\newcommand{\TRA}{TRA\xspace}
\newcommand{\LOC}{LOC\xspace}
\newcommand{\RT}{\textit{QoM-agnostic}\xspace}
\newcommand{\RTH}{\textit{QoM-aware}\xspace}
\newcommand{\etal}{\textit{et al.}\xspace}
\newcommand{\eg}{\textit{e.g.}\xspace}

\newcommand{\ie}{\textit{i.e.}\xspace}

\begin{document}

\title{HetSched: Quality-of-Mission Aware Scheduling for Autonomous Vehicle SoCs
}

\author{
\IEEEauthorblockN{Aporva Amarnath}\footnote{\label{note}Parts of this work was done when the authors were graduate students at University of Michigan}
\IEEEauthorblockA{
\textit{IBM Research}\\
Yorktown Heights, NY \\
aporva.amarnath@ibm.com}
\and
\IEEEauthorblockN{Subhankar Pal}
\IEEEauthorblockA{
\textit{IBM Research}\\
Yorktown Heights, NY \\
subhankar.pal@ibm.com}
\and
\IEEEauthorblockN{Hiwot Kassa}
\IEEEauthorblockA{
\textit{University of Michigan}\\
Ann Arbor, MI \\
hiwot@umich.edu}
\and
\IEEEauthorblockN{Augusto Vega}
\IEEEauthorblockA{
\textit{IBM Research}\\
Yorktown Heights, NY \\
ajvega@us.ibm.com}
\and
\IEEEauthorblockN{Alper Buyuktosunoglu}
\IEEEauthorblockA{
\textit{IBM Research}\\
Yorktown Heights, NY \\
alperb@us.ibm.com}
\and
\IEEEauthorblockN{~~~~~~~Hubertus Franke}
\IEEEauthorblockA{
~~~~~~~\textit{IBM Research}\\
~~~~~~~Yorktown Heights, NY \\
~~~~~~~frankeh@us.ibm.com}
\and
\IEEEauthorblockN{John-David Wellman}
\IEEEauthorblockA{
\textit{IBM Research}\\
Yorktown Heights, NY \\
wellman@us.ibm.com}
\and
\IEEEauthorblockN{Ronald Dreslinski}
\IEEEauthorblockA{
\textit{University of Michigan}\\
Ann Arbor, MI \\
rdreslin@umich.edu}
\and
\IEEEauthorblockN{Pradip Bose}
\IEEEauthorblockA{
\textit{IBM Research}\\
Yorktown Heights, NY \\
pbose@us.ibm.com}
}



\maketitle

\begin{abstract}
Systems-on-Chips (SoCs) that power autonomous vehicles (AVs) must meet stringent performance and safety requirements prior to deployment.
With increasing complexity in AV applications, the system needs to meet stringent real-time demands of multiple safety-critical applications simultaneously.
A typical AV-SoC is a heterogeneous multiprocessor consisting of accelerators supported by general-purpose cores.
Such heterogeneity, while needed for power-performance efficiency, complicates the art of task (process) scheduling.

In this paper, we demonstrate that hardware heterogeneity impacts the scheduler's effectiveness and that optimizing for only the real-time aspect of applications is not sufficient in AVs. 
Therefore, a more holistic approach is required --- one that considers global \textit{Quality-of-Mission} (QoM) metrics, as defined in the paper. 
We then propose \SCHED, a multi-step scheduler that leverages dynamic runtime information about the underlying heterogeneous hardware platform, along with the applications' real-time constraints and the task traffic in the system to optimize overall mission performance. 
\SCHED proposes two scheduling policies: \MSstat and \MSdyn and scheduling optimizations like task pruning, hybrid heterogeneous ranking and rank update. 
\SCHED improves overall mission performance on average by 4.6$\times$, 2.6$\times$ and 2.6$\times$ when compared against CPATH, ADS and  2lvl-EDF (state-of-the-art real-time schedulers built for heterogeneous systems), respectively, and achieves an average of 53.3\% higher hardware utilization, while meeting 100\% critical deadlines for real-world applications of autonomous driving and aerial vehicles. Furthermore, when used as part of an SoC design space exploration loop, in comparison to the prior schedulers, \SCHED reduces the number of processing elements required by an SoC to safely complete AV's missions by 35\% on average while achieving 2.7$\times$ lower energy-mission time product.
\end{abstract}
\setlength{\textfloatsep}{0.1cm}
\setlength{\floatsep}{0.1cm}

\section{Introduction}
\label{sec:introduction}

With the growing prominence of fully autonomous vehicles (ground, aerial, surface and underwater), major investments are being made into developing applications to make these vehicles efficient and safe. In order to ensure functionally correct \emph{and} safe operation, the complexity of state-of-the-art full-stack hardware-software platforms for autonomous vehicles (AVs) has progressively increased over the last decade --- specifically in the form of highly-heterogeneous hardware systems driven by highly-heterogeneous software applications. The resulting ``nominal'' hardware-software platform for AVs consists of domain-specific systems-on-chips (DSSoCs) with multiple acceleration engines specifically catering to ultra-efficient execution of application kernels for perception, planning, control, and communication. Examples of these platforms include NVIDIA's DRIVE AGX~\cite{driveagx} and Tesla's Full Self-Driving (FSD) chip~\cite{teslafsd}.
In this context, the existing work focuses mostly on: (i) optimal SoC platforms for AVs to comply with stipulated performance, efficiency and resiliency metrics, and (ii) the development of AV applications that can meet the increasing functionality and safety requirements for autonomy. 
Little attention has been paid to the aspect of process scheduling for AV applications on heterogeneous DSSoCs. The \emph{de facto} approach relies on schedulers developed for:

\begin{enumerate}[leftmargin=*]
\setlength\itemsep{0pt}

\item Heterogeneous systems (traditional distributed systems and not SoCs, therefore low variation in execution time across processing elements)~\cite{heft,cpath}.

\item Real-time constrained applications~\cite{edf,priority}.

\item Real-time constrained applications based on the ones listed in 1) above~\cite{rheft,wu2017bsf_edf,xie2017adaptive}.
\end{enumerate}

However, none of these schedulers use dynamic runtime information from the system to efficiently schedule real-time applications on heterogeneous SoCs. 
Moreover, these schedulers operate in a greedy manner, trying to meet the real-time requirements of individual processes or applications without any consideration of a \textit{global} objective function, defined as the Quality-of-Mission (QoM) metric for AV applications. We define the QoM as a composite metric encompassing both the mission performance, the fraction of mission completed \textit{safely}, \ie while meeting the real-time and safety constraints and the energy consumed.

To demonstrate the value of considering the QoM, we examine the following two schedulers:
\begin{enumerate}[leftmargin=*]
\item \textbf{Quality-of-Mission \textit{Agnostic} Approach.} The tasks of an AV application, represented as directed-acyclic graphs (DAG), are statically scheduled using the earliest-finish-time and lower-bound approach detailed in \cite{xie2017adaptive}. During execution, based on the \textit{safety-criticality level} of a DAG (a specification that indicates how critical the timely execution of a task is for safe AV operation), the tasks are re-ordered to execute on the fastest processing element (PE).

\item \textbf{Quality-of-Mission \textit{Aware} Approach.} Tasks are ranked taking into account the temporal density of safety-critical DAGs in the system,  real-time constraints (deadline and criticality-level) {along with} their heterogeneous execution profiles and dynamic runtime information. This allows the scheduler to make ``smarter'' scheduling decisions across PEs in highly-heterogeneous SoCs while navigating through dynamic environments, as we propose in this paper.
\end{enumerate}

\begin{figure}[tb]
\centering
  \includegraphics[width=1\columnwidth]{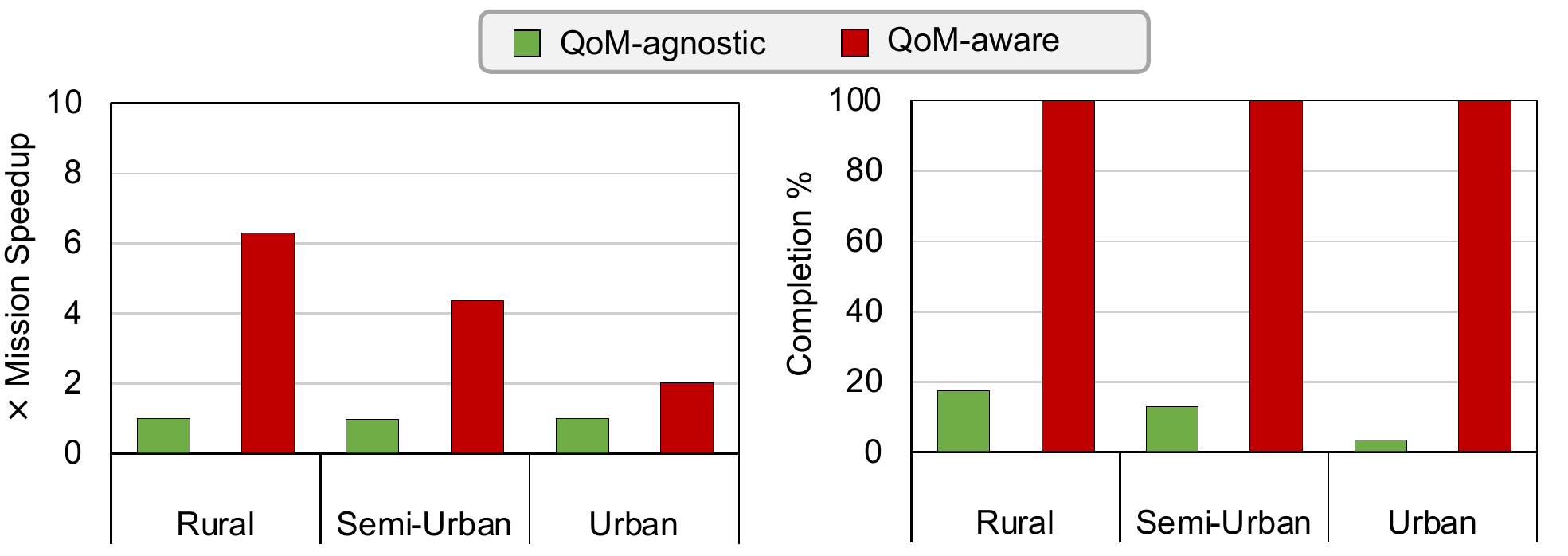}
  \vspace{-\baselineskip}
  \caption{\textit{Left-axis:} Mission speedup of Quality-of-Mission aware scheduler (\RTH) over the Quality-of-Mission agnostic scheduler (\RT) and \textit{Right-axis:} Mission completed by \RT and \RTH while operating at the speed achieved by \RTH, when evaluated under three congestion scenarios: rural, semi-urban and urban.}
\label{fig.motivation_chart}
\end{figure}

Figure~\ref{fig.motivation_chart} presents the evaluation of these scheduler variants under progressively more congested navigation conditions (rural, semi-urban and urban). The figure evaluates the schedulers on two metrics: (i) the overall mission time while meeting all real-time constraints and (ii) the percentage of mission completed safely when the AV is operating at the maximum safe speed achievable among the two schedulers. The schedulers are evaluated on a simulated platform with eight general-purpose cores, two GPUs, and one fixed-function hardware accelerator. To complete a \textit{mission} (e.g. safely navigate from a starting location ``A'' to a destination ``B''), the SoC executes a series of \textit{applications}, composed of \textit{tasks} (or kernels or processes). Examples of AV tasks include perception, planning and control.

Figure~\ref{fig.motivation_chart} reveals that scheduling decisions can drastically affect the safe speed of the AV, and consequently its overall mission completion time across varying congestion levels.
The \RTH scheduler outperforms the \RT scheduler in terms of mission time by 6.3$\times$, 4.4$\times$ and 2.0$\times$ for the rural, semi-urban and urban scenarios, respectively. Similarly, when \RT is operated at the safe speed achieved by \RTH, it is able to complete only up to 17\% of the mission before encountering a hazard.
This motivates the notion that a holistic approach that is aware of the heterogeneity in hardware and applications along with dynamic run-time information can help make better scheduling decisions even in a highly congested urban-like scenario. Moreover, using this information, the scheduler can stall or prune less-critical applications in favor of more critical ones, or prioritize the execution of a given task on an accelerator over other tasks which may need to use the same accelerator -- the approach followed by \RTH. The key observation here is that \textbf{real-time constrained execution of AV applications without accounting for hardware heterogeneity and dynamic runtime information does not necessarily translate to the best overall mission performance (\textit{e.g.} mission time).}

\red{In this work, we propose a \RTH scheduler called \SCHED that has been developed as a part of DARPA's DSSoC program that seeks to develop hardware-software co-design to build efficient DSSoCs. One aspect of the program concentrates on the development of intelligent scheduling for heterogeneous SoCs.}
\SCHED is a multi-level scheduler that leverages the synergy between the underlying heterogeneous hardware platform and the applications' runtime characteristics to satisfy the growing throughput demand of AVs, while meeting the specified real-time and safety constraints. 
Specifically, \SCHED proposes two scheduling policies: \MSstat and \MSdyn, in addition to scheduling optimizations, such as \textit{task-traffic reduction, hybrid heterogeneous ranking and rank update}. 
The first step in the operation of \SCHED involves profiling the application tasks across \textit{all} the possible PEs in the SoC\footnote{Note that offline application profiling is a common approach across most of the schedulers considered in this work.}. 
This information is then used by \SCHED to guide its scheduling decisions. 
\begin{figure}[b]
\centering
  \includegraphics[width=1\columnwidth]{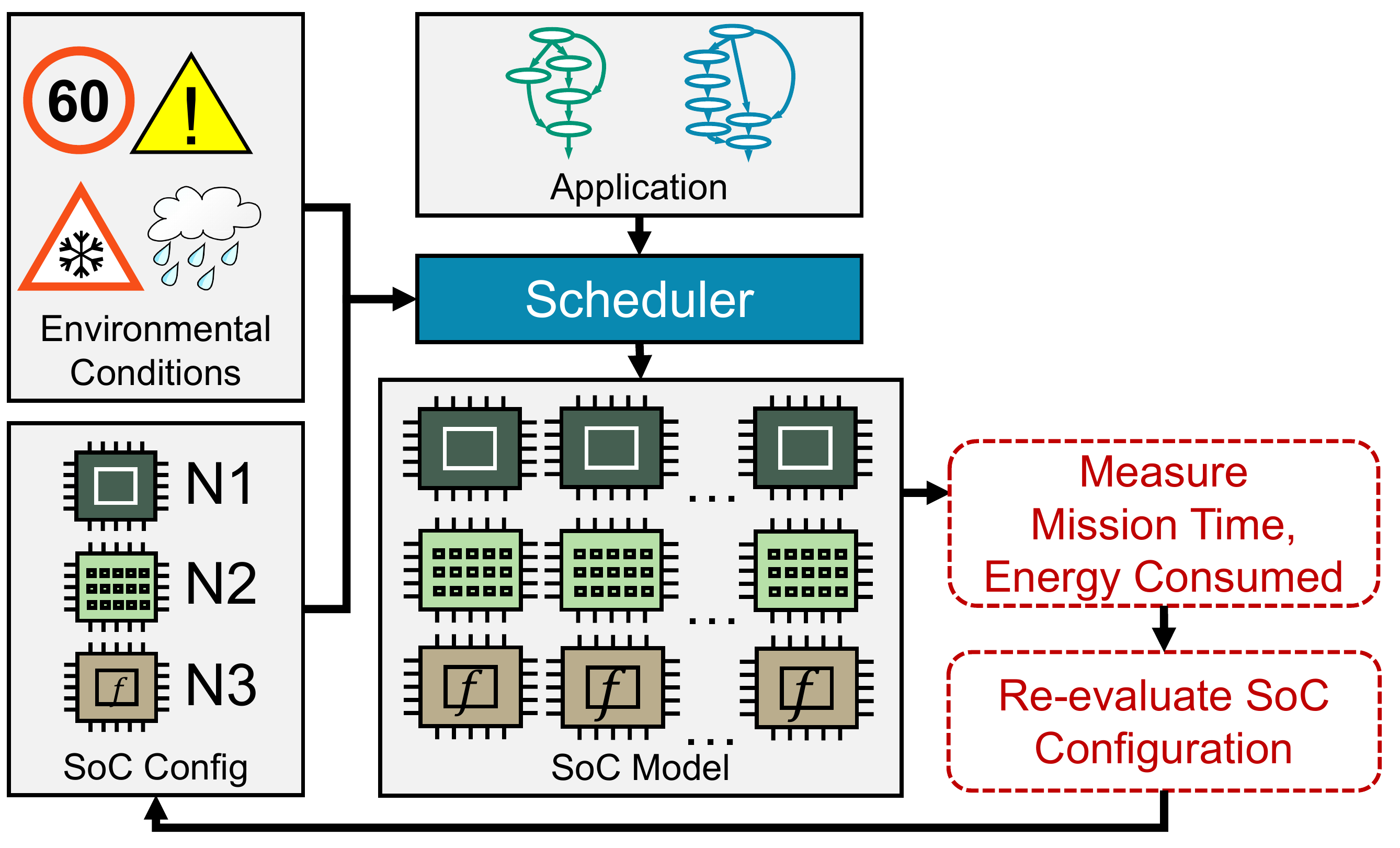}
  \caption{\red{Use of \SCHED to optimize SoC design for AVs while considering QoM metrics. Use of an efficient scheduler in this optimization loop helps reduce the compute requirements of an SoC while improving the mission performance and energy consumption.}}
\label{fig:soc_dse}
\end{figure}

\SCHED also uses safety criticality information provided by the application, which is \textit{key} to comply with safety specifications. 
Runtime information gathered from hardware monitors in the SoC are used by \SCHED during operation to keep track of real-time deadlines of the application, and to estimate data movement costs, wait-times of ready tasks, slack available for a DAG, and power consumed by a completed task. 
These monitors include, but are not limited to, the status of PEs ({available}/{busy}), estimated completion time for tasks running on busy PEs, and the execution profile of completed tasks. 
\red{Moreover, efficient design space exploration of various processing elements (PEs) in the SoC can be achieved by using \SCHED to optimize for the mission time and energy consumption for AV applications constrained by different environmental conditions as shown in Figure~\ref{fig:soc_dse}.}

Specifically, the contributions of this paper are as follows.

\begin{enumerate}[leftmargin=*]
\setlength\itemsep{0pt}

\item We demonstrate that hardware heterogeneity along with the application's runtime information is key in determining the scheduler's effectiveness while unveiling new opportunities for smarter task scheduling.

\item We propose \SCHED, a multi-level scheduler that follows a holistic approach to optimize the \textit{Quality-of-Mission} (QoM), while meeting real-time safety constraints in autonomous vehicles. The scheduler exploits the highly-heterogeneous nature of the underlying SoC and dynamic run-time information (like maximum/minimum slack available and task wait times) to make better scheduling decisions.

\item We introduce optimizations such as \textit{task pruning}, \textit{hybrid-ranking} and \textit{rank update} built upon two \SCHED policies (\MSstat and \MSdyn) that result in a performance improvement of 3.2$\times$ (average) for real-world AV applications when compared with state-of-the-art schedulers.

\item We show significant reductions in mission time, SoC energy and area using \SCHED in a design space exploration (DSE) loop for SoC design. We show that \SCHED reduces the accelerator resource requirement of an efficient SoC to safely complete AV missions by 35\% (on average), compared to state-of-the-art schedulers~\cite{wu2017bsf_edf, xie2017adaptive}.
\end{enumerate} 
\section{Background and Motivation}
\label{background_and_motivation}

\subsection{Autonomous Vehicle Applications}

To achieve high levels of safety, reliability and precision, AV applications are constituted of highly heterogeneous tasks that can be divided into three types based on their function: \textit{perception}, \textit{planning} and \textit{control}~\cite{pendleton2017perception}. Through perception tasks, AVs sense the environment and perform obstacle detection, localization and classification to determine further action. Planning tasks are implemented to make decisions in order to achieve the vehicle's goals such as reaching a destination or searching an unknown location while ensuring safety and mission quality. Lastly, control tasks such as traction control, 
acceleration, braking, steering, and lane keeping are executed to follow the planned actions. We describe the characteristics of and models for AV applications in the following paragraphs.

\subsubsection{Heterogeneity in Execution Time}
Task execution times can vary by orders of magnitude across PEs on a heterogeneous AV SoC~\cite{lin2018architectural}. 
In our experiments, we observed this variation to be up to 300$\times$ (Section~\ref{evaluation}).
Therefore, to make heterogeneity-aware scheduling decisions, we rely on an offline timing profile  for each task that can be stored on-chip along with the application binary. A task's timing profile comprises both the intra-PE {execution cost} as well as the inter-PE {data movement cost} of inputs/outputs for all eligible PEs. The execution cost and the data-movement cost also depend on the number of other tasks in the system and memory contention.

\subsubsection{Application Model}
\label{subsubsec:application_model}
For AVs, all applications and their conforming tasks are fixed at runtime, \ie the addition of a new application (with its offline timing profile) would be provided as a software update to the AV.
Based on the type of AV and its mission, these tasks are executed according to a fixed control-flow graph (CFG), where edges in the graph are dynamically decided based on the inputs and decisions made during runtime. \red{We derive directed-acyclic graphs (DAGs) as subgraphs from these CFGs that are statically known, although the arrival and execution of these DAGs are dynamic and determined during vehicle operation. These dynamically arriving static DAGs constitute the input to the scheduler.}
\blue{A DAG contains nodes and edges. We map a task in a CFG to a node in the DAG and dependencies between tasks as edges. Note that a task is a independent unit of work that can execute when its data/control dependencies are resolved.}
DAGs are generated using compiler techniques for extraction of basic blocks from the CFG.

\subsubsection{Safety Criticality Level of Applications}

Depending on the AV's operating environment, each iteration of the CFG can execute at a different safety-criticality level. For autonomous driving applications, ISO~26262 identifies four Automotive Safety Integrity Levels (ASIL): A, B, C, and D~\cite{asil2018levels}. ASIL-A represents the lowest criticality (\ie operations which can result in no injuries) and D represents the highest criticality (\ie operations that can result in the highest degree of automotive hazard). Similarly, unmanned AV tasks have criticality levels assigned based on the Design Assurance Level (DAL)~\cite{rtca1992software}. For the safe and reliable operation of AVs, it is absolutely necessary to comply with these criticality levels. In this paper, we consider DAGs to belong to two criticality levels:

\noindent\textbf{Non-Critical DAGs}: those with criticality 1 (\texttt{crit=1}) that arrive periodically to the system. They are equivalent to ASIL A, \eg object recognition on a blind-spot camera while traveling straight on a single-lane road.
    
\noindent\textbf{Critical DAGs}: those with criticality 2 (\texttt{crit=2}), that are classified as critical in two ways:
\begin{itemize}[leftmargin=*]
    \item {\textit{Promoted DAGs}: If no \texttt{crit=1} DAG meets its deadline within a time period $T_\text{crit}$, then the scheduler can promote it to \texttt{crit=2} in order to provide redundancy and avoid potential hazards in the AV operation, \eg a path-planning operation that uses GPS to calibrate the location of the AV while it is moving along a straight line.}
    \item \textit{Highly-Critical DAGs:} DAGs that 
    represent applications of ASIL levels which can result in a clear safety hazard (B, C, and D) would be \texttt{crit=2} DAGs, \eg forward-camera perception of a stop sign during forward motion.
\end{itemize}

\noindent For safe AV operation, DAGs with \texttt{crit=2} have to be executed within specific \textit{hard deadlines} in order to avoid potential hazards. DAGs with \texttt{crit=1} have \textit{firm deadlines}, \ie if executed within their deadlines they could help improve the mission. Otherwise, the output of the DAG is not useful.

\subsection{Congestion in Environmental Conditions}
The safety and resilience of AVs are of prime importance due to the toll they can have on human lives and infrastructure~\cite{uberaccident, vlasic2016self}. Hence, the assessment of AV systems operating in varying dynamic scenarios is crucial~\cite{monitor, faultinjection, handsoff}. 
The congestion of an environment is determined by the temporal density of \texttt{crit=2} DAGs encountered during execution. This can be influenced by conditions like the weather, traffic and terrain. \textit{E.g.} in the case of autonomous driving, the vehicle might encounter several crosswalks while driving from point A to point B in an urban area. In this case, the AV passing each crosswalk could be accompanied by the arrival of a \texttt{crit=2} DAG. Therefore, the more congested the environment (e.g. more crosswalks), the higher the number of \texttt{crit=2} DAGs that the AV will have to execute. 
In this work, we consider three congestion scenarios: \textit{rural}, \textit{semi-urban} and \textit{urban}, however we are not limited by this classification.

\subsection{Application Deadline and Speed of the AV}
The AV speed determines the rate at which DAGs arrive at the scheduler. 
Each DAG is also associated with a \textit{real-time deadline}, determined by the speed of the vehicle (the faster the AV, the tighter the deadline), congestion in the environment and the application criticality.
{Hence, the AV speed is directly proportional to the rate of DAG arrival and inversely proportional to the deadline}. The maximum arrival rate at which the AV meets 100\% critical deadlines is considered equivalent to its \textit{maximum safe speed} for a given congestion scenario.

\subsection{Quality-of-Mission (QoM) Metrics}\label{subsec:qom}
Various figures of merit can be used to measure an AV’s mission quality. In this paper, we use universal metrics that can be applied to all AVs, similar to the ones adopted in \cite{boroujerdian2018mavbench}.  We choose the following QoM metrics to evaluate our scheduler for varying congestion scenarios:
\begin{itemize}[leftmargin=*]
\setlength\itemsep{0pt}
    \item \textit{Mission time} to complete the objective of the mission, \eg navigation time from location $A$ to location $B$, while complying with safety requirements of meeting deadlines for all DAGs with \texttt{crit=2}.
    \item \textit{Fraction (or \%) of mission completed} at a given speed before missing the first \texttt{crit=2} DAG deadline. For example, if the best scheduler is able to complete a mission safely while operating at a maximum speed of $S$, then this metric for the scheduler being evaluated is calculated as the \% of total critical DAGs of the mission an AV running at speed $S$ is able to complete before it fails to meet the deadline of a critical task leading to a hazard.
    \item \textit{Energy consumed} by the AV SoC for mission completion.
\end{itemize}

\subsection{Domain-Specific Systems-on-Chips}

High heterogeneity in AV applications, real-time constraints, and the demand to process multiple critical applications call for the use of highly heterogeneous systems. These SoC platforms consist of multiple processing elements (PEs) with different performance and efficiency characteristics; namely, CPUs, GPUs, accelerators, \textit{etc.}~\cite{teslafsd, nvidiaOrin, mobileye, aerialSoC}. Heterogeneous SoCs accelerate the execution of a task by providing increased computational capabilities, reduced data movement cost between PEs, and reduced need to offload computation to cloud servers (or other vehicles, in case of connected vehicle systems~\cite{hashim2019application,8500627}), in addition to higher energy efficiency.

\begin{figure}[t]
\centering
  \includegraphics[width=1\columnwidth]{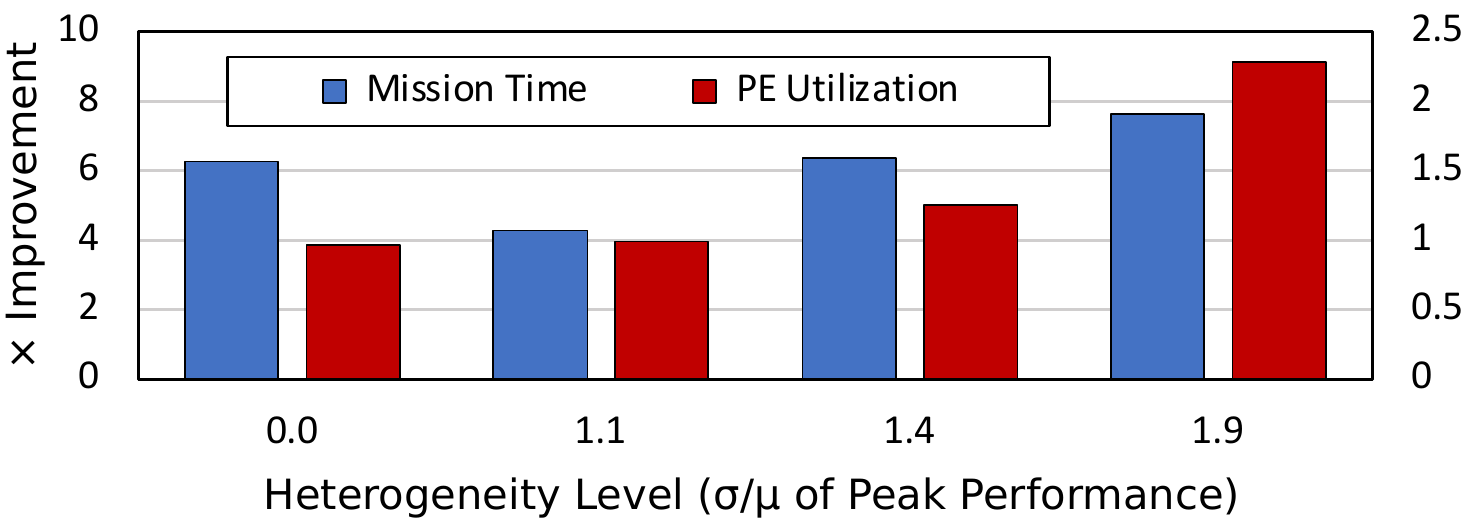}
  \caption{Effect of SoC heterogeneity, calculated as the coefficient of variation in PE peak performance, on mission speedup and PE utilization of a Quality-of-Mission-aware scheduler (\RTH) over a Quality-of-Mission agnostic (\RT). We increase heterogeneity in the system by considering varying number of PEs in the SoC.}
\label{fig:hetero}
\end{figure}

The heterogeneity in PEs (Table~\ref{tab:exec_times}) results in new challenges and opportunities when allocating on-chip resources or making task scheduling decisions. To illustrate the need for global schedulers that are aware of the heterogeneity in an AV SoC, we compare the quality-of-mission agnostic scheduler (\RT) with the quality-of-mission aware scheduler (\RTH), as described in Section~\ref{sec:introduction}, for different hardware configurations. We use the \textit{coefficient of variation}~\cite{coefficient_of_variation}
of the PE's peak performance as a proxy for the heterogeneity level in the SoC. Figure~\ref{fig:hetero} shows that as we increase heterogeneity (by diversifying the PEs), \RTH is able to improve performance by up to 7.6$\times$ over \RT. By leveraging this heterogeneity information, \RTH is also able to improve PE utilization by up to 2.2$\times$ over \RT. The takeaway is that synergistic exploitation of the underlying hardware, the application's real-time requirements (deadline and criticality) and dynamic runtime information can significantly improve mission time and hardware utilization.

\subsection{State-of-the-Art Real-Time \textit{and} Heterogeneous Schedulers}

\label{subsec:sota}

The processing time of any task comprises of four components: the transfer of input data to the PE that will execute the task (\textit{data movement time}), the time required to make the scheduling decision (\textit{scheduling decision time}), the time spent while the task is waiting to be executed on the scheduled PE (\textit{waiting time}), and the time to execute the task on the scheduled PE (\textit{execution time}). In order to minimize the mission time of an AV, it is critical to reduce all four components. While data movement and execution time are significantly reduced by the use of heterogeneous SoCs, all four components are also highly dependent on the scheduling algorithm. Prior schedulers developed for heterogeneous data-center architectures do help curtail the processing time, but are neither \textit{hetero-aware} (do not efficiently schedule tasks with high-variation in timing profile on an SoC) nor do they optimize for stringent real-time and safety constraints. Furthermore, schedulers developed for real-time-constrained applications are not flexible enough to provide the best QoM metrics or efficiently utilize the underlying hardware. Table~\ref{tab:prior_schedulers} provides a comparison of this work with prior art and briefly discusses them here.

\noindent\textbf{CPATH}~\cite{cpath} is a scheduler that prioritizes tasks in the DAG based on a bottom-cost longest-path approach and submits high priority tasks to fast cores and low priority tasks to slow cores with work-stealing enabled. \red{CPATH aims to optimize the response time of a single DAG. When applied to a multi-DAG application with real-time constraints, it fails to meet deadlines at higher arrival rates of DAGs. In contrast, our work targets to meet deadlines in safety-critical multi-DAG scenarios.}

\noindent\textbf{2lvl-EDF} schedules tasks with the earliest deadline on the earliest finish time PE, as described in \cite{wu2017bsf_edf}. \red{However, it neither considers safety constraints of tasks nor the variation in the timing profile of tasks on the heterogeneous SoC with respect to deadlines.}

\noindent\textbf{ADS} schedules ranked DAGs based on \cite{heft} and dynamically prioritizes tasks with higher criticality levels, as described in \cite{xie2017adaptive}. \red{However, ADS neither predicts when to prune non-critical tasks, nor is hetero-aware. \SCHED is able to outperform this policy by pruning non-critical tasks, which is further enhanced by \SCHED's hybrid heterogeneous ranking optimization.}

None of these prior schedulers operate efficiently on highly-heterogeneous SoCs while optimizing for the real-time requirements of the application and improving overall AV mission performance. Our work targets to fill this void.

\begin{table}[t]
\centering
\caption{Real-Time and Heterogeneous Schedulers.}

\label{tab:prior_schedulers}
\resizebox{1\columnwidth}{!}{
\begin{tabular}{|c|c|c|c|c|c|}
\hline
\textbf{Prior Art} & \textbf{Input} & \begin{tabular}[c]{@{}c@{}}\textbf{Ranking Type,}\\\textbf{Parameters}\end{tabular} & \begin{tabular}[c]{@{}c@{}}\textbf{Hetero-}\\\textbf{Aware}\end{tabular} & \begin{tabular}[c]{@{}c@{}}\textbf{QoM-}\\\textbf{Aware}\end{tabular} \\ \hline

CPATH~\cite{cpath} & \begin{tabular}[c]{@{}c@{}}Single\\Dynamic DAG\end{tabular} & \begin{tabular}[c]{@{}c@{}}Dynamic\\ Critical path length\end{tabular} & \textcolor{red}{$\times$} & \textcolor{red}{$\times$} \\ \hline

2lvl-EDF\cite{wu2017bsf_edf}& \begin{tabular}[c]{@{}c@{}}Multiple\\Static DAGs\end{tabular} & \begin{tabular}[c]{@{}c@{}}Dynamic\\ Earliest deadline first \end{tabular} & \textcolor{red}{$\times$} & \textcolor{red}{$\times$}\\ \hline

ADS~\cite{xie2017adaptive} & \begin{tabular}[c]{@{}c@{}}Multiple\\Static DAGs\end{tabular} & \begin{tabular}[c]{@{}c@{}}Static, Earliest finish time,\\ criticality\end{tabular} & \textcolor{red}{$\times$} & \textcolor{red}{$\times$}\\ \hline

\textbf{\SCHED} & \begin{tabular}[c]{@{}c@{}}Multiple\\Static DAGs\end{tabular} & \begin{tabular}[c]{@{}c@{}}Dynamic, Deadline, \\PE variation, criticality\end{tabular} & \textcolor{blue}{\Checkmark} & \textcolor{blue}{\Checkmark} \\ \hline
\end{tabular}
}

\end{table}

\section{Mission- \& Heterogeneity-Aware Scheduling}
\label{scheduling_heuristics}

\blue{\SCHED is a multi-level scheduler that exploits the heterogeneous nature of domain-specific SoCs to improve QoM and PE utilization for AV applications. Specifically, \SCHED consists of two levels: \META and \TASK.  \META translates the mission and application (DAG) level information to tasks, while \TASK performs the actual task-to-PE assignment and resource management.} The two layers communicate using a set of data structures: a \textit{ready queue}, a \textit{completed queue} and a \textit{prune list}.

As depicted in Figure~\ref{fig:overview}, when DAGs arrive for execution, \META tracks task dependencies, prioritizes ready tasks based on rank, and performs pruning of non-critical tasks.
\TASK receives ready tasks from \META, updates tasks' ranks, populates the prune list, assigns tasks to PEs, and sends information of completed tasks to \META.

\subsection{Meta-Sched and Task-Sched Operations}
\label{subsec:meta_tsched_op}

This section describes \META and \TASK operation and introduces the various scheduling features in \SCHED. Some key terms are defined in Table~\ref{tab:rank_parameters}.

\subsubsection{Dependency Tracking}
\blue{\META processes DAGs to find \textit{ready} tasks. A task is determined as \textit{ready} when all its parent nodes in the DAG have completed execution; \ie it has no incomplete dependencies (incoming edges) in the DAG. Therefore, when a task completes execution on a PE, \META resolves edges to the children tasks and marks those with no remaining dependencies as ready.} 

\subsubsection{Task Prioritization}\label{subsubsec:ranking}
Each application (DAG) arriving at \META has an associated deadline and criticality. Moreover, these DAGs have varying structures in terms of the number of tasks, types of tasks (execution profile) and dependencies (edges). Therefore, to make an informed task scheduling decision (\ie considering the real-time constraints of the DAG and mission performance), \META assists \TASK by assigning ranks to ready tasks and ordering them. The rank encodes DAG- and mission-level information as it is determined using the deadline of the parent DAG (the DAG to which the task belongs), criticality and structure, and the task's execution profile. A task's \textit{rank} is calculated as: 

\begin{equation}
Rank = \dfrac{Criticality}{Slack}; \;\;  \mathit{Eff\_Slack} = SD - EET
\label{eq:slack}
\end{equation}

\begin{table}[t]
\centering
\caption{Description of parameters used in \SCHED.}

\label{tab:rank_parameters}
\resizebox{1\columnwidth}{!}{
\footnotesize
\begin{tabular}{|l|l|}
\hline
\textbf{Abbreviation} & \textbf{Parameter Description}                                               \\ \hline
WCET / BCET/ ACET                   & Task's worst/best/avg.-case execution time across all PEs \\
\hline
EET                    & Task's estimated execution time                 \\
\hline
PT                     & Sum of all WCET of tasks in the path            \\
\hline
CPT                     & Sum of all WCET of tasks in the critical path            \\
\hline
\multirow{2}{*}{CPST}  & Sum of all WCET of tasks in the segment of the path \\
                       & that intersects the critical path\\
\hline
\multirow{2}{*}{NCPST} & Sum of all WCET of tasks in the segment of the path \\
                       &  that does not intersect the critical path\\
\hline
SD / SDR / SR          & Task's sub-deadline / sub-deadline ratio / slack ratio                             \\
\hline
\end{tabular}
}
\end{table}

\noindent where, $Criticality$ is the criticality of the task determined by that of the parent DAG, and $Eff\_Slack$ is the task's effective slack calculated by \META as the task's sub-deadline ($SD$) minus its estimated execution time ($EET$). \blue{Therefore, tasks with higher criticality and smaller slack to their deadline are given higher priority.} We explore multiple rank assignment policies based on the way $SD$ and $EET$ are computed. We use the parent DAG's structure and the task's execution profile to determine $SD$, \ie for the path within the parent DAG on which the task lies, we find the worst-case execution time ($WCET$) of the path and of the task. The $WCET$ of a task is the time required execute the task on the slowest PE. Therefore, by using $WCET$ to calculate $SD$, \SCHED allows for the tasks to be scheduled on any available PE in the system, whereas using average or best case execution time can bias the scheduler's decision towards faster PEs.
Depending on the way $SD$ is calculated, \SCHED policies are classified as follows:

\begin{figure}[tb]
\centering
  \includegraphics[width=1\columnwidth]{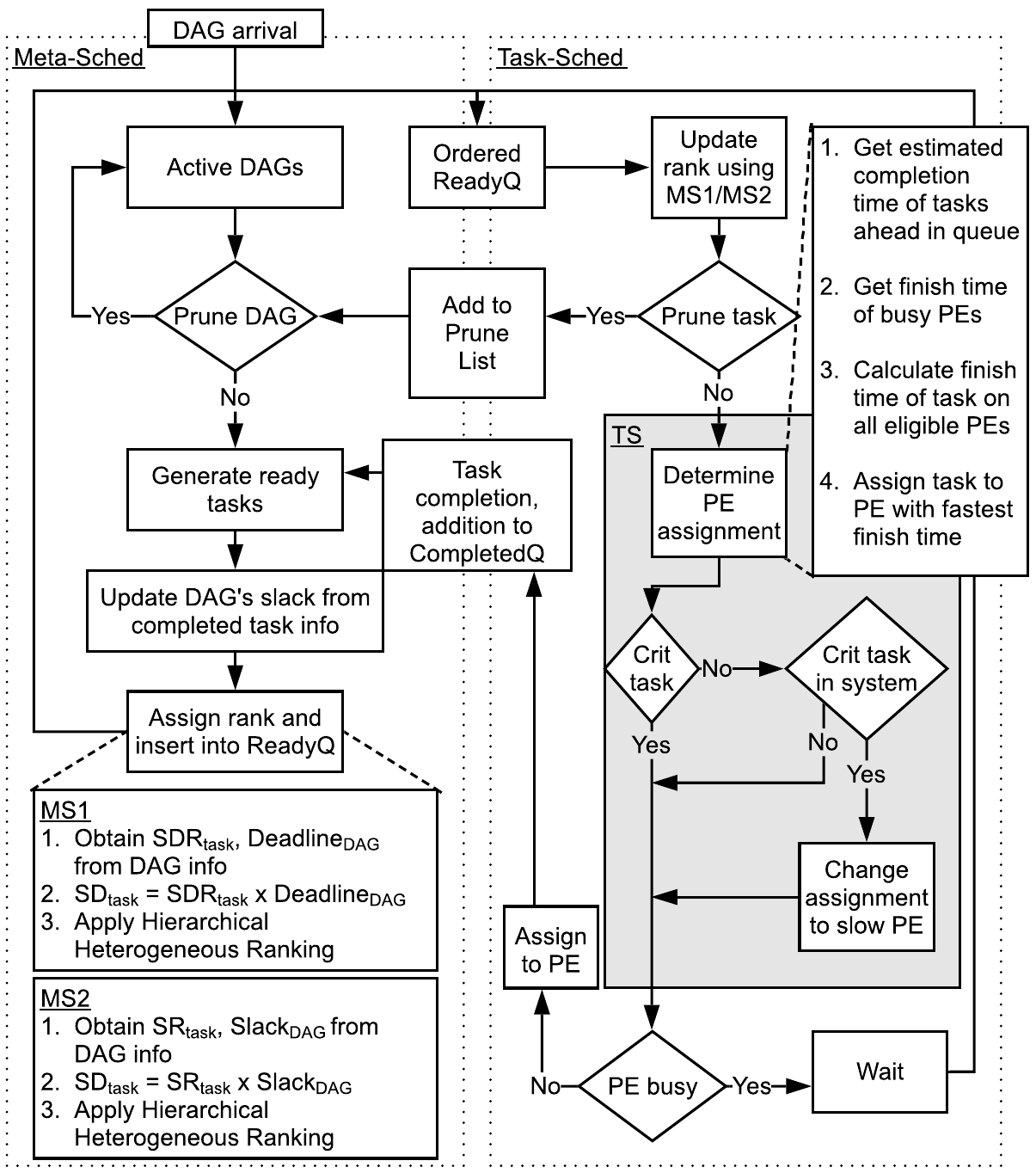}

  \caption{\red{\SCHED operations showing \META (mission \& DAG processor) and \TASK (task scheduler \& hardware manager), and their synchronization using the ready and completed queues, and prune list.}}
\label{fig:overview}

\end{figure}
\noindent\textbf{\MSstat:} 
In the \MSstat policy, we determine $SD$ as a weighted-ratio of the DAG's deadline ($Deadline_{DAG}$). This weighted-ratio, called the task's sub-deadline ratio ($SDR$), is calculated as the task's $WCET$ relative to its path's execution time. Since each DAG and timing profile of the tasks in it are statically known, $SDR$s can be calculated offline and stored along with the timing profile of the DAG. For a given DAG, the path time ($PT$) is calculated as the sum of the $WCET$s of the tasks in that path. The critical path time ($CPT$) is the one with the largest $PT$. \red{A task's $SDR$ and $SD$ calculation are based on the path of the DAG on which it lies:}

\begin{itemize}[leftmargin=*]
\item \red{If the task lies on the critical path or on a path that does not intersect with the critical path, SDR is calculated as the ratio of the $WCET$ of the task to the path time $PT$:}
    \begin{equation}
    SDR = \dfrac{WCET}{PT}; \;\;  SD = SDR \times {Deadline}_{DAG} 
\label{eq:sd_ms1}
    \end{equation}

\item If the task lies on a path which intersects with the critical path, then the path is divided into two segments, the critical path segment (taking a critical-path segment time, $CPST$) and the non-critical path segment (with non-critical-path segment time, $NCPST$). For tasks on the $NCPS$, we first calculate the deadline allocated to the $NCPS$, ${Deadline}_{NCPS}$, as the ${Deadline}_{DAG}$ minus ${Deadline}_{CPS}$. Using Equation \ref{eq:sd_ms1}, 

    \begin{equation}
        {Deadline}_{CPS} = \dfrac{CPST}{CPT}\times{Deadline}_{DAG}
        \label{eq:deadline_cps}
    \end{equation}
    
    \begin{equation}
        {Deadline}_{NCPS} = {Deadline}_{DAG} - {Deadline}_{CPS}
        \label{eq:leftover}
    \end{equation}
    
For tasks on the $NCPS$, $SDR$ and $SD$ are calculated similarly, using a path time of $NCPST$ and deadline of ${Deadline}_{NCPS}$ as:

    \begin{equation}
    SDR = \dfrac{WCET}{NCPST}; \;\;  SD = SDR \times {Deadline}_{NCPS}
    \label{eq:sd_ncps}
    \end{equation}

\end{itemize} 

\begin{figure}[t]
\centering
  \includegraphics[width=1\columnwidth]{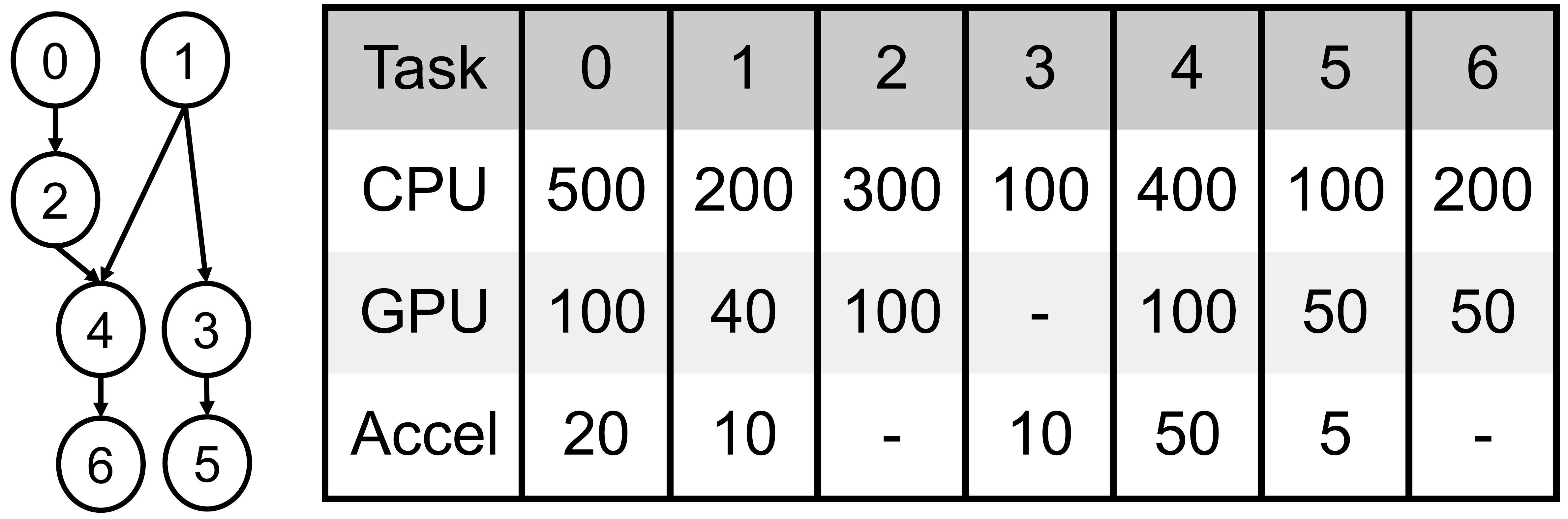}

  \caption{\textit{Left:} A 7-task DAG containing three paths: $P_{0}$, $P_{1}$ and $P_{2}$. $P_{0}$ contains tasks 0, 2, 4 and 6; $P_{1}$ contains tasks 1, 4 and 6; and $P_{2}$ contains tasks 1, 3 and 5. \textit{Right:} Timing profile for each task on three types of PEs: CPU, GPU and an accelerator. Using the timing profile, we determine that $P_{0}$ is the critical path, $P_{1}$ intersects the critical path, and $P_{2}$ is independent of it.}

\label{fig:sdr_sr}
\end{figure}

If a given task's sub-deadline $SD$ can be calculated using several of the above methods, we pessimistically assign it the smallest of the computed $SD$ values.
To illustrate with an example, Figure~\ref{fig:sdr_sr} shows a small, 7-task DAG. Let path $P_{0}$, the path consisting of tasks 0, 2, 4 and 6 be the critical path. $P_{1}$ contains tasks 1, 4 and 6 and $P_{2}$ is composed of tasks 1, 3 and 5. While $P_{1}$ intersects the critical path, $P_{2}$ does not. Therefore, $SDs$ for tasks on $P_{0}$ and $P_{2}$, are calculated using Equation~\ref{eq:sd_ms1}. Since the $SD$ for task 1 on the NCPS of $P_{1}$, $SD$ can also be calculated using Equation~\ref{eq:sd_ncps}, we assign it the lower value of the two.

\setlength{\textfloatsep}{0.1cm}
\setlength{\floatsep}{0.1cm}

\noindent\textbf{\MSdyn:} In this policy, we assign the task's sub-deadline based on a dynamic metric of the DAG. Specifically, $SD$ is calculated using the DAG's available slack ($Slack_{DAG}$): the deadline \textit{remaining} for a DAG during execution, when that ready task's rank is being calculated, as opposed to \texttt{\MSstat} that uses a static distribution of the DAG's deadline to calculate the task's $SD$. Therefore, \MSdyn accounts for tasks in the DAG that might have exceeded their sub-deadlines. \MSdyn adjusts the $SD$ of ready tasks based on the available slack of the DAG by calculating a task's $WCET$ relative to the execution time of tasks \textit{remaining} in the task's path:

    \begin{equation}
    SR = \dfrac{WCET}{\sum WCET_{i}}; \;\;  SD = SR \times Slack_{DAG}. 
    \label{eq:sd_ms2}
    \end{equation}

\noindent where, $SR$ is the task's slack ratio and $WCET_{i}$ is the $WCET$ of each remaining task $i$ that lies on the same path as the task, including the task itself.
If a task lies on multiple paths, then lowest $SR$ calculated across all paths is selected. 
For the DAG in Figure~\ref{fig:sdr_sr}, we calculate $SD$ and $SR$ of the tasks using Equation~\ref{eq:sd_ms2}. Since task 1 lies on two paths, $P_{1}$ and $P_{2}$, its $SR$ is calculated twice and we choose it to be smaller of the two.

\textit{Homogeneous Ranking}: In this ranking scheme, we calculate $rank_{hom}$ as \textit{Crit}/{$\mathit{Eff\_Slack}$}, 
where the $\mathit{Eff\_Slack}$ is determined by using the $WCET$ of the task as the $EET$ in Equation~\ref{eq:slack}. Therefore, this ranking policy prioritizes critical tasks that have earlier deadlines over non-critical tasks that have later deadlines. 

\begin{figure}[tb]
\centering

  \includegraphics[width=0.8\columnwidth]{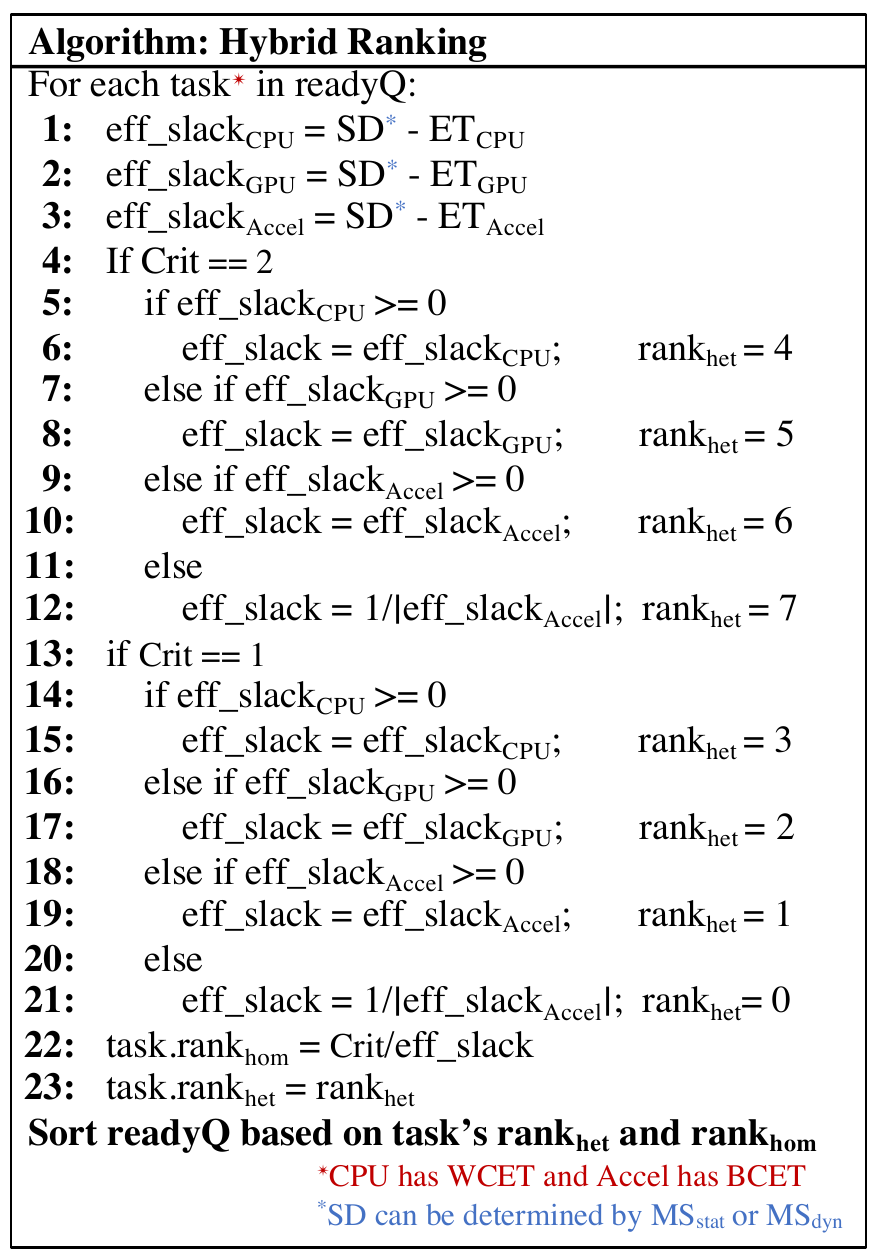}

  \caption{Algorithm for hybrid ranking of tasks using both $rank_{hom}$ and $rank_{het}$ heuristics.}

\label{fig:algorithm}
\end{figure}

\textit{Heterogeneous Ranking: }This ranking approach, $rank_{het}$, improves upon $rank_{hom}$ by accounting for the variation in the execution time of different PEs on the SoC, as well as by adopting contrasting approaches based on the task's \textit{Crit}.
The goal of the scheme is to allow for critical tasks to be prioritized on fast PEs and non-critical tasks to be run on slow PEs without blocking the fast PEs. For this, we first calculate $\mathit{Eff\_Slack}$ for each PE type that the task can execute on, as shown in Figure~\ref{fig:algorithm} using Equation~~\ref{eq:slack},~\ref{eq:sd_ms1} and~\ref{eq:sd_ncps}. We then use the task's \textit{Crit} and $\mathit{Eff\_Slack}$ of each PE type to calculate $rank_{het}$. $rank_{het}$ serves two purposes: (i) prioritization of non-critical tasks that can meet their deadlines on more than one type of PE over those that can meet their deadlines only when executed on a fast PE; and (ii) prioritization of critical tasks with earlier deadlines over those with later deadlines, while considering the PEs that it can execute on.

\textit{Hybrid Ranking}:
To differentiate between tasks of the same $rank_{het}$ in the heterogeneous scheme, we additionally, introduce the scheme of ``hybrid ranking'' that prioritizes tasks based on both $rank_{hom}$ and $rank_{het}$. However, unlike the homogeneous ranking scheme that uses $WCET$ for $\mathit{Eff\_Slack}$ to calculate  $rank_{hom}$, we assign $\mathit{Eff\_Slack}$ based on the PEs that the task can execute on to meet its deadline. This method, shown in Figure~\ref{fig:algorithm}, is similar to how $rank_{het}$ is determined. Thus, allowing for tasks having the same $rank_{het}$ to be prioritized using $rank_{hom}$ (earliest deadline first). Hybrid ranking also allows for tasks that cannot be executed on all PE types to be prioritized using $rank_{hom}$ and $rank_{het}$ for eligible PE types. Note that this case is not depicted in Figure~\ref{fig:algorithm}.

\subsubsection{Rank Update}
\TASK receives ordered ready tasks from \META. Before the task assignment, the ranks of the tasks waiting in the ready queue are updated to subtract the time elapsed since previous update from the current effective slack ($Eff\_Slack$). Updating the ready tasks' ranks can also help in finding non-critical tasks that might not meet their deadline, which can then be considered candidates for pruning, reducing the overall system task traffic. \TASK passes these tasks to \META for potentially pruning their parent DAGs using the \textit{prune list}.

\subsubsection{Task Assignment}\label{subsubsec:task_assignment}

\TASK uses a task assignment policy to assign ordered ready tasks in the ready queue to eligible PEs (\ie PEs that can execute it). Furthermore, once a task completes execution, it is pushed into the completed queue along with information about the PE on which it was executed and the timestamp at which it completed execution. We introduce a non-blocking task assignment policy, called \texttt{TS},
that schedules a task in the ready queue to the PE that will result in the earliest estimated finish time for the task, factoring in the execution time of the task, current busy status of the PE and all tasks ahead of this task in the ready queue that are potentially scheduled to the same PE as shown in Figure~\ref{fig:overview}. \texttt{TS} chooses the task to be scheduled using a non-blocking task assignment policy within a window of size $w$, thus searching $w$ tasks past the head of the queue that could potentially be waiting for the earliest estimated finish time PE to become available. \texttt{TS} is also aware of the timing profile and criticality of each task. Therefore, if the task is non-critical and critical tasks are present in the system, \texttt{TS} can  improve utilization by scheduling this task on available slow PEs (Figure~\ref{fig:overview}).

\subsubsection{Completed Task Information}
Once a task completes execution, \TASK pushes it into the completed queue along with information about the PE on which it was executed and the timestamp at which it completed execution. This information is used by \META for dependency tracking and to obtain the data movement cost of children tasks.

\subsubsection{Deadline Tracking and Task Pruning}
\META can elect to prune DAGs,
\ie not execute them at all/any further, thus eliminating non-critical tasks that will not meet their deadlines in order to reduce traffic in the system. After the execution of each task, when \META searches the existing DAGs for ready tasks, it also calculates the estimated slack available for each DAG, assuming that ready tasks can execute at their best-case execution time ($BCET$). If the estimated slack available is negative and the DAG has \texttt{crit=1}, the entire DAG is pruned. \blue{\META also prunes DAGs based on the tasks in the prune list, identified during the rank update process.} Note that in Algorithm~\ref{fig:algorithm}, \META prunes DAGs that have $rank\_type$ 0 or 1, \textit{only if} there are critical DAGs in the system.

\subsection{Summary of Cumulative \SCHED Features}
\label{subsec:summary}

In summary, \SCHED introduces the following scheduling policies and optimizations.

    \subsubsection{Task-Sched policy, \texttt{{TS}}} A non-blocking scheduler, that schedules ready tasks to the PE with the fastest projected finish time. \blue{\texttt{TS} also schedules non-critical tasks to the slowest PEs, if critical tasks are present in the system.}
    \subsubsection{Two-level scheduling policies} With pruning of non-critical tasks estimated to miss their deadlines. \blue{These scheduling policies prioritize ready tasks based on their rank, calculated using the criticality, sub-deadline and estimated execution time of the task}, and use \texttt{TS} for their \TASK. \\
        \noindent\texttt{\textbf{\MSstat}} \blue{determines the sub-deadline of a task statically from the parent DAG's deadline. \texttt{\MSstat} performs best when the deadlines of DAGs are significantly large, \ie when DAGs complete execution with large remaining slack and all tasks are able to complete within their assigned sub-deadlines.} \\ 
        \noindent\texttt{\textbf{\MSdyn}} \blue{ uses the task's parent DAG's available slack, computed during execution, to dynamically calculate the sub-deadline of the task. Due to this ability to adapt to changes in execution time of tasks, including missing task deadlines, \texttt{\MSdyn} performs best for stringent DAG deadlines.}
    \subsubsection{Scheduling optimizations for \texttt{\MSstat} and \texttt{\MSdyn}}\textcolor{white}{:}\\
    \textbf{Heterogeneous ranking} accounts for the variation in execution time of a task on the heterogeneous SoC using dynamic calculation of the rank.\\
    \textbf{Hybrid ranking} uses the effective slack of the tasks along with hetero-ranking to improve overall mission performance by incorporating the state of the system. \\
    \textbf{Rank Update and Task Pruning} \blue{revises the task ranks to incorporate time waiting in the ready queue or when critical tasks are encountered. This feature also identifies non-critical tasks that will not meet their deadlines and should be pruned.}

\section{Experimental Methodology}
\label{system_description}

\subsection{Hardware Description}\label{subsec:hw_desc}

We first profile (offline) a set of AV kernels on an NVIDIA TX1 board, which is representative of an SoC used in real-world AV systems. 
This information is then used to simulate a heterogeneous SoC with multiple PEs. 
\red{We assume that the simulated SoC has variants of the Arm Cortex-A57 CPU and the NVIDIA Maxwell GPUs with 256 CUDA cores, and fixed-function accelerators for certain tasks. 
We consider three systems, named $Sys_A$, $Sys_B$ and $Sys_C$, the hardware descriptions of which are shown in Table~\ref{tab:hw_desc}.} We also consider a unified memory (shared physical address space) between the PEs in the simulated SoC. 
\SCHED, however, is not limited to this specific choice.

\begin{table}[t]
\centering
\caption{System Description of SoC for Workload Evaluation.}

\label{tab:hw_desc}
\resizebox{1\columnwidth}{!}{
\footnotesize
\begin{tabular}{|l|c|l|}
\hline
\textbf{Workload} & \textbf{System} & \textbf{Description}                                               \\ \hline
Synthetic     & $Sys_A$              & \begin{tabular}[c]{@{}l@{}}8 single-core Arm Cortex-A57 CPUs\\2 NVIDIA Maxwell GPUs\\1 CNN/FFT accelerator~\cite{chen2016eyeriss, seok20110}\end{tabular} \\
\hline
& $Sys_B$                 & \begin{tabular}[c]{@{}l@{}}8 single-core Arm Cortex-A57 CPUs\\2 NVIDIA Maxwell GPU\\1 tracking accelerator~\cite{lin2018architectural}\\1 localization accelerator~\cite{lin2018architectural}\\1 detection accelerator~\cite{lin2018architectural}\end{tabular}  \\
\cline{2-3}
\begin{tabular}[c]{@{}l@{}}Real-World\\\\\\\\\\\end{tabular} & \red{$Sys_C$}                & \begin{tabular}[c]{@{}l@{}}$N_{C}$$^{\dagger}$ Single-core Arm Cortex-A57 CPUs\\$N_{G}$$^{\dagger}$ NVIDIA Maxwell GPUs\\$N_{traA}$$^{\dagger}$ tracking accelerators~\cite{lin2018architectural}\\$N_{locA}$$^{\dagger}$ localization accelerators~\cite{lin2018architectural}\\$N_{detA}$$^{\dagger}$ detection accelerators~\cite{lin2018architectural}\end{tabular}  \\
\hline

\hline
\end{tabular}
}

$^{\dagger}$$N_{C}$,$N_{G}$,$N_{traA}$,$N_{locA}$,$N_{detA}$ are determined using design space exploration

\end{table}

\subsection{Application Task Profile}

\subsubsection{Synthetic Application Tasks}
Our synthetic applications are comprised of three types of tasks: 2D Fast Fourier Transform (fft), 2D convolution (conv) and Viterbi decoding (decoder), taken from the Mini-ERA benchmark suite~\cite{miniera}, which simulates a simplified AV with minimal environmental conditions. 
For fft, we use  FFTW3~\cite{FFTW05}  (CPU) and cuFFT (GPU). For conv, we use  Arm Compute Library~\cite{armcomputlib} powered by Neon SIMD extensions (CPU), and cuDNN~5.1 (GPU). We also obtain timing profiles of fft and conv for the accelerators in~\cite{chen2016eyeriss, seok20110}. Finally, for {decoder}, we use the GNURadio Viterbi function (CPU)~\cite{miniera} and a PyCUDA implementation~\cite{pyterbi} (GPU). 

\subsubsection{Real-World Application Tasks}
We consider two real-world AV benchmarks; ADSuite and MAVBench. 

\noindent \textbf{ADSuite}~\cite{lin2018architectural} provides an autonomous driving application comprised of kernels like object detection (\DET), object tracking (\TRA), localization (\LOC), mission planning and motion planning. For \DET, we use YOLOv3~\cite{redmon2018yolov3}, a DNN-based detection algorithm, on a series of 7 images derived from the VOC dataset~\cite{Everingham15}.
We use the Tiny-YOLOv3 pre-trained set of weights, which is much faster and lightweight, but less accurate compared to the regular YOLO model.
For \TRA, we use GOTURN~\cite{held2016learning}, a DNN-based single object tracking algorithm, on a series of 14 videos in the ALOV++ dataset~\cite{smeulders2013visual}.
For LOC, we use ORB-SLAM~\cite{mur2015orb}, a highly-ranked vehicle localization algorithm, on 3 sequences from the KITTI datasets~\cite{geiger2013vision}.
Further, for our GPU evaluation, we adopt the ORB-SLAM implementation in~\cite{orbSlamGpu}, where the hot paths are rewritten using CUDA. We also obtain timing profile of \DET, \TRA and \LOC on their respective accelerators from \cite{lin2018architectural}.
For motion and mission planning, we use the \texttt{op\_local\_planner} and  \texttt{op\_global\_planner}~\cite{darweesh2017open} kernels in Autoware~\cite{kato2015open}. The fusion kernel combines the coordinates of the objects being tracked with the AV location. It has a small latency, for which we only consider CPU execution.

\noindent \textbf{MAVBench}~\cite{boroujerdian2018mavbench} provides a set of computational kernels that form the building blocks of many aerial vehicle applications. For some of the kernels, we use different algorithms than the ones in~\cite{boroujerdian2018mavbench}, in order to better exploit the heterogeneity in our hardware. Specifically, for the perception, tracking and localization kernels, we reuse our ADSuite implementations. For occupancy map generation, we use OctoMap~\cite{hornung13auro} and GPU-Voxels~\cite{hermann2014unified} for the CPU and GPU implementations, respectively. OctoMap performs 3D occupancy grid mapping, and GPU-Voxels is a CUDA-based library for robotics planning and monitoring tasks. We generate a map composed of 200$\times$200$\times$200 voxels. Point cloud generation and collision check consume 10--1000$\times$ lower latency in comparison to other kernels (Table I in~\cite{boroujerdian2018mavbench}), and so we only employ CPU implementations for these. 
For the shortest-path planners, we use the CPU-based parallel RRT (pRRT)~\cite{devaurs2011parallelizing} implementation in the Open Motion Planning Library (OMPL)~\cite{sucan2012open}, on the ``Cubicle'' benchmark. For the GPU implementation, we use a Poisson-disk samples based GPU algorithm~\cite{park2013realtime}. For frontier exploration, we use the RRT ROS package that implements a multi-robot RRT-based map exploration algorithm~\cite{umari2017autonomous}, on the ``single\_simulated\_house'' scenario. Finally, we consider only CPU execution for path tracking as it has a low latency.

We consider two applications from MAVBench in this work, namely Package Delivery, where an aerial AV navigates through an obstacle-filled environment to reach its destination, deliver a package and return back to its origin, and 3D Mapping, that instructs the aerial AV to build a 3D map of an unknown polygonal environment specified by its boundaries.

\subsubsection{Data Movement Cost}
{To build a realistic evaluation model, we profiled the data movement time across each pair of PE types in the system.
We consider the cost for data movement \textit{within} a PE to be zero, \ie if two dependent tasks execute on the same PE, there is no additional overhead.
Data movement between a CPU core and a GPU is assumed to be equivalent to the time to flush the parent tasks' output from the CPU's caches into main memory, thereby allowing the GPU to load the input data of the child task from the same memory location, since the CPU and GPU share the same physical address space. 

The data movement cost from a GPU to a different PE is encapsulated in the timing profile of the task on the GPU. For an accelerator, we consider the data movement cost from/to the accelerator to be the direct-memory access (DMA) transfer cost, since many accelerator designs have their own local memory. We derived empirical data movement cost for the CPU and GPU by profiling the TX1 board, and use DMA transfer rates from published specifications~\cite{arm2010axi} for the accelerators.}

\subsubsection{Resource Contention}\label{sssec:contention}
The execution of a task on a given PE in a realistic SoC is naturally impacted by the volume of parallel tasks executing across the PEs, \ie the contention that the given PE faces due to resources shared with other PEs in the system, such as conflicts at the interconnect network, reduced effective cache capacity if there is no data sharing, \textit{etc.} We used gem5~\cite{binkert2006m5} to simulate a system with $N+1$ PEs, in order to model the contention due to $N$ PEs. The cache hierarchy and memory are modeled after the TX1 SoC. We constructed an analytical model of the contention cost across the problem sizes used in the evaluated applications, which we used while simulating the benchmarks on \SCHED.

\subsection{Energy and Power Model}\label{subsec:energy_power_model}
\noindent\textbf{Power and Energy Estimations.} We profiled the average power consumption of each task by measuring the power consumed on the VDD rails of the CPU and GPU of the TX1 board. For the accelerators, we use the estimated power values reported in the prior work (Section~\ref{subsec:hw_desc}). To compute the end-to-end energy of the SoC with \SCHED, we sum the energy consumed for each task on a PE.

\noindent\textbf{Dynamic Voltage-Frequency Scaling (DVFS).}
We apply DVFS techniques, similar to those in~\cite{pillai2001real,aydin2004power}, on the PEs to recuperate low utilization using a fraction of the available slack considering each task's sub-deadline, which is dependent on the scheduling policy -- \MSstat or \MSdyn (Section~\ref{subsec:meta_tsched_op}). Prior to a task being scheduled onto a PE, the target clock frequency is selected based on: (i) the estimated slack, (ii) the current clock frequency, and (iii) a factor $f_{\text{slack}}$ that defines the fraction of slack to be recuperated. Note that na\"ively applying DVFS on the full estimated slack ($f_{\text{slack}}$=$1$) \textit{may} lead to deadline violations and consequently failure of the mission, \eg if a task $T_i$, running on a GPU, is slowed down too much, a critical task $T_{i+1}$, that was formerly waiting on this GPU, could be scheduled onto a slower core. 
For the purposes of this work, we pessimistically apply a static value for $f_{\text{slack}}$ across all DAGs and tasks in a DAG. In the real world, DVFS governors can be integrated within the scheduler to dynamically select $f_{\text{slack}}$ per task based on the current PE utilization.

We enable DVFS only for the CPU and GPUs, since we observed that DVFS for even small values of $f_{\text{slack}}$ on accelerators leads to low energy savings and mission failures. We use voltage ($V_{DD}$) and clock frequency ($f$) points from the embedded DVFS tables in the TX1 to obtain scaling factors for the PE voltage and clock. We further assume  the execution time of a task to scale linearly with the PE's operating frequency, and  the power to scale as $V_{DD}^{2.5}$ using DVFS~\cite{ChipBasics}.

\begin{table*}[t]
\centering
\caption{Timing and power profile of evaluated kernels on each PE-type.}
\resizebox{1.2\columnwidth}{!}{
\begin{tabular}{|l||l|l|l|l|l|l|l|}
\hline
\multicolumn{1}{|c||}{\multirow{2}{*}{\textbf{Suite}}}                         & \multicolumn{1}{c|}{\multirow{2}{*}{\textbf{Task Type}}}                      & \multicolumn{3}{c|}{\textbf{Execution Time}}                                 & \multicolumn{3}{c|}{\textbf{Power (mW)}}                                                                   \\ \cline{3-8} 
\multicolumn{1}{|c||}{}                                                        & \multicolumn{1}{c|}{}                                                         & \multicolumn{1}{c|}{\textbf{CPU}} & \multicolumn{1}{c|}{\textbf{GPU}} & \multicolumn{1}{c|}{\textbf{ASIC}} & \multicolumn{1}{c|}{\textbf{CPU}} & \multicolumn{1}{c|}{\textbf{GPU}} & \multicolumn{1}{c|}{\textbf{ASIC}} \\ \hline\hline
\multirow{3}{*}{Mini-ERA}                                                     & 2D Convolution                                                                & 583$^*$                           & 349$^*$                           & 180$^*$                            & 634                               & 2225                              & 445                                \\ \cline{2-8} 
                                                                              & Viterbi Decoder                                                               & 1021$^*$                          & 20$^*$                            & -                                  & 864                               & 1228                              & -                                  \\ \cline{2-8} 
                                                                              & 2D FFT                                                                        & 3193$^*$                          & 97$^*$                            & 4$^*$                             & 1036                              & 6364                              & 4                               \\ \hline
\multirow{3}{*}{\begin{tabular}[c]{@{}l@{}}ADSuite /\\ MAVBench\end{tabular}} & Object Detection                                                              & 3531$^{\dagger}$                  & 156$^{\dagger}$                   & 96$^{\dagger}$                     & 3654                              & 467                               & 28                                 \\ \cline{2-8} 
                                                                              & Object Tracking                                                               & 1825$^{\dagger}$                  & 17$^{\dagger}$                    & 2$^{\dagger}$                     & 5600                              & 12790                             & 590                                \\ \cline{2-8} 
                                                                              & Localization                                                                  & 165$^{\dagger}$                   & 95$^{\dagger}$                    & 10$^{\dagger}$                     & 6133                              & 4457                              & 22                                 \\ \cline{2-8} & {Mission Planning}    & \white{000}1$^{\dagger}$          & -                                 & -                         & 3534              & -                 & -             \\ \cline{2-8}
& {Motion Planning}     & \white{000}8$^{\dagger}$          & -                                 & -                         & 4222              & -                 & -             \\  \hline
ADSuite                                                                       & Fusion                                                                        & 0.1$^{\dagger}$                   & -                                 & -                                  & 505                               & -                                 & -                                  \\ \hline
\multirow{5}{*}{MAVBench}                                                     & \begin{tabular}[c]{@{}l@{}}Occupancy Map Gen\\ + Point Cloud Gen\end{tabular} & 976$^{\dagger}$                   & 761$^{\dagger}$                   & -                                  & 2995                              & 3533                              & -                                  \\ \cline{2-8} 
                                                                              & Shortest Path Planner                                                         & 1005$^{\dagger}$                  & 379$^{\dagger}$                   & -                                  & 3302                              & 3533                              & -                                  \\ \cline{2-8} 
                                                                              & Collision Check                                                               & 1$^{\dagger}$                     & -                                 & -                                  & 500                               & -                                 & -                                  \\ \cline{2-8} 
                                                                              & Path Tracking                                                                 & 1$^{\dagger}$                     & -                                 & -                                  & 501                               &                                   &                                    \\ \cline{2-8} 
                                                                              & Frontier Exploration                                                          & 397$^{\dagger}$                   & -                                 & -                                  & 5980                              & -                                 & -                                  \\ \hline
\end{tabular}
}

\hspace{150pt}*in micro-seconds~~~~~~ $\dagger$in milli-seconds

\label{tab:exec_times}  
\end{table*}

\subsection{Trace Generation }
\subsubsection{Synthetic DAG Traces}
\blue{In order to evaluate the generality of \SCHED, we generate synthetic traces of DAGs arriving at the scheduler that represent applications executed by AVs for varying congestion scenarios. Each entry of the trace consists of the arrival time, type, criticality and deadline of the DAG. }

\blue{The type of DAG is determined by the composition of the tasks and dependencies between them. We generate different types of DAGs consisting of 5 to 10 tasks of three types of tasks (\texttt{fft}, \texttt{conv} and \texttt{decoder}). A DAG can have a criticality level of 1 or 2, and the fraction of \texttt{crit=2} DAGs in the trace reflects the congestion in the environment. Each DAG's deadline is set as the critical path time ($CPT$). We generate 1,000 DAG traces for the three congestion scenarios (urban, semi-urban and rural) with \texttt{crit=2} DAG fractions of 50\%, 20\%, and 10\% for urban, semi-urban and rural, respectively. We then evaluate these traces at varying DAG arrival rates (AV speeds) to determine the best QoM metrics.}

\subsubsection{Real-World Application Traces} 
ADSuite and MAVBench have kernel components that make up the end-to-end applications’ CFGs. To generate DAGs, we take the CFGs of both ADSuite and MAVBench and study scenarios that can lead to the execution of different sets of kernels. Such scenarios can arise from the vehicle changing route and leading to the execution of the mission planning kernels, and so on. Using the set of kernels executed in the CFG for a particular scenario, we generate DAGs with varying deadlines and criticalities.
For ADSuite, as described in \cite{lin2018architectural}, we set the deadline of each critical path task to be 100~ms. As no such information is available for MAVBench, each DAG’s deadline is set as the $CPT$. We generate 1,000 DAG traces and choose the same \texttt{crit=2} DAG fractions for the three congestion scenarios.

\subsection{Simulation Platform}\label{subsec:stomp}

To explore multiple AV workloads and flexible SoC configurations that are not offered by a fixed real-world system, \SCHED is implemented on the STOMP open-source scheduler evaluation platform~\cite{2007.14371}. STOMP is a queue-based discrete-event simulator used for early-stage evaluation of task scheduling mechanisms in heterogeneous platforms. 
We augment STOMP to accept DAG-based inputs, while using the underlying queue-based simulator to schedule ready tasks. We also added real-time parameters, such as deadlines and safety criticalities. We realistically model the simulated SoC by providing STOMP with the power and timing profile of tasks obtained from the TX1 platform (Table~\ref{tab:exec_times}). \red{STOMP also provides the flexibility to add a deviation to the execution time to account for contention on shared resources like memory and buses, which we add as described in Section~\ref{sssec:contention}.}

\section{Evaluation}\label{evaluation}
We evaluate \SCHED for synthetic and real-world traces derived from AV benchmark suites Mini-ERA~\cite{miniera}, ADSuite~\cite{lin2018architectural} and MAVBench~\cite{boroujerdian2018mavbench}. We also incrementally evaluate our proposed features and compare with state-of-the-art schedulers, namely  CPATH~\cite{cpath}, ADS~\cite{xie2017adaptive} and 2lvl-EDF~\cite{wu2017bsf_edf} in terms of QoM metrics (Section~\ref{subsec:qom}) and overall PE utilization.

The offline timing profiles generated for key kernels of both the synthetic and real-world applications for representative input data sizes are shown in Table~\ref{tab:exec_times}.

\subsection{\SCHED Optimizations}

\begin{figure}[b]
\centering
  \includegraphics[width=1\columnwidth]{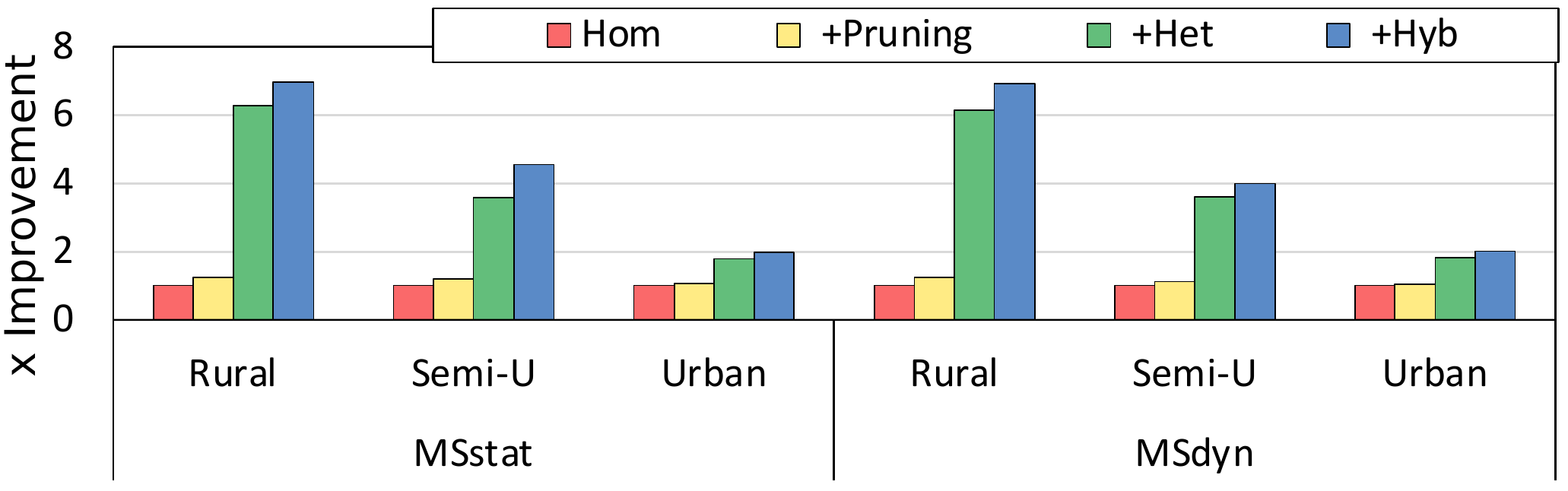}

  \caption{Improvements in mission time on $Sys_A$ using \SCHED with task pruning (\texttt{Pruning}), heterogeneous-ranking (\texttt{Het}) and hybrid-ranking (\texttt{Hyb}) over \SCHED with homogeneous-ranking (\texttt{Hom}) for different congestion scenarios.}
\label{fig:feat_eval}
\end{figure}

\SCHED introduces two-level schedulers, that use \texttt{TS} as \TASK and \texttt{\MSstat} and \texttt{\MSdyn} along with optimizations for \META. We evaluate the optimizations (pruning, hetero-rank ordering and hybrid ranking) to analyze the benefits from each, under varying congestion levels. These features are evaluated with the rank update optimization.

\subsubsection{Task Pruning}
Mission time improvement of \texttt{\MSstat} and \texttt{\MSdyn} along with \texttt{Pruning} over their respective homogeneous ranking based scheduler \texttt{Hom} for different environments is shown in Figure~\ref{fig:feat_eval}. Both \texttt{\MSstat} and \texttt{\MSdyn} achieve speedups of up to 1.3$\times$ over \texttt{Hom}.

\subsubsection{Hetero- and Hybrid Ranking}
Employing hetero-ranking (\texttt{Het}) for \texttt{\MSstat} and \texttt{\MSdyn} helps achieve a reduction in mission time by up to 6.3$\times$ and 6.1$\times$ for \texttt{\MSstat} and \texttt{\MSdyn}, as shown in Figure~\ref{fig:feat_eval}. This increased reduction is a combined benefit of \texttt{Pruning} and \texttt{Het}, as hetero-ranking identifies more candidate tasks that can be pruned. Applying the hybrid ranking (\texttt{Hyb}) feature over \texttt{Het} helps to improve mission performance by 2.0--7.0$\times$ for \texttt{\MSstat} and \texttt{\MSdyn}. 

\subsection{Scheduler Evaluation for Real-World Applications}\label{subsec:eval}

For each of the real-world applications, we compare \SCHED against prior baseline schedulers, namely CPATH, ADS and 2lvl-EDF, in terms of QoM metrics and PE utilization. Note that the energy reported here for all systems accounts for total power of PEs during task execution and static power when the PE is idle.

\subsubsection{Metrics Comparison}

\begin{figure}[t]
\centering
  \includegraphics[width=1\columnwidth]{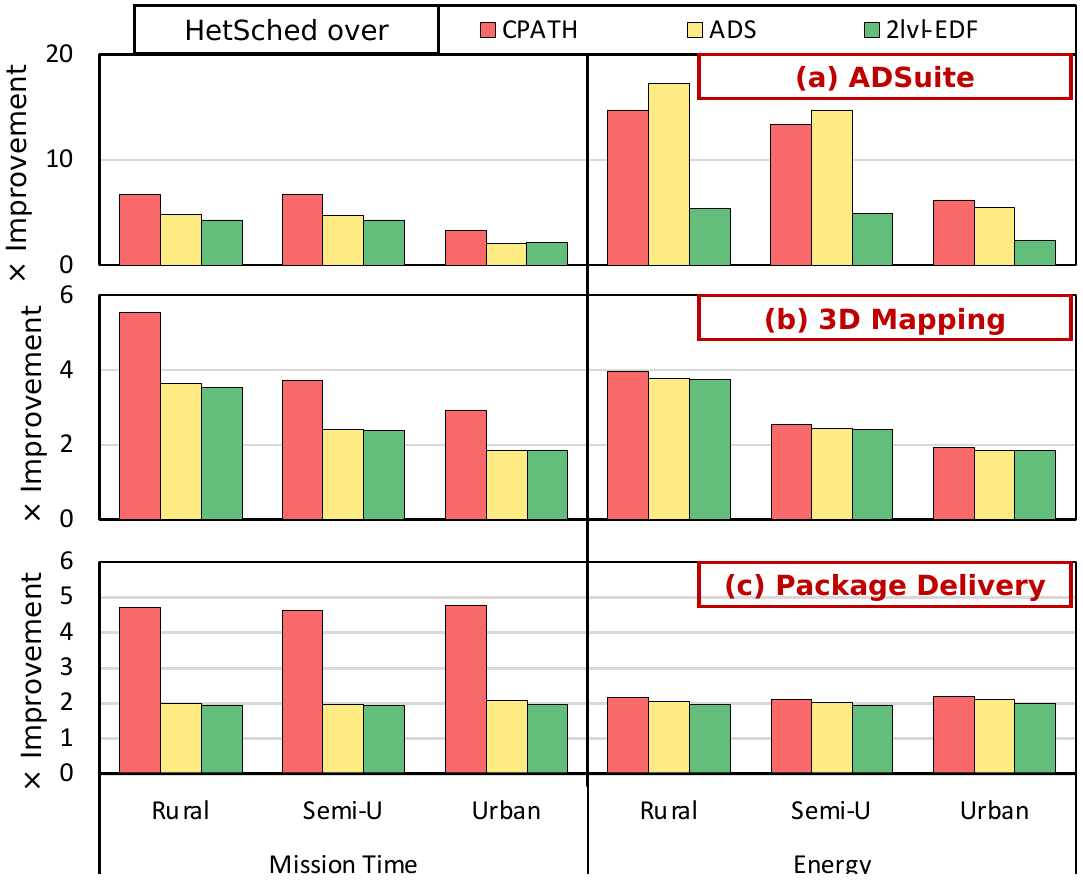}

  \caption{\red{Comparison of mission time (\textit{left}) and energy (\textit{right}) on $Sys_B$ of \SCHED against prior-work schedulers for (a) ADSuite, (b) 3D Mapping, and (c) Package Delivery.}}
\label{fig:real_world}
\end{figure}

For ADSuite (Figure~\ref{fig:real_world}(a)), \SCHED enabled with task pruning, hybrid ranking and rank update achieves a 2.1-6.8$\times$ improvement in mission time \red{for $Sys_B$} over the baseline schedulers. In terms of \% mission completed, for $Sys_B$ (not shown), the state-of-the-art schedulers complete just 38\%, 5\% and 7\% of the mission at the maximum safe speed of \SCHED for the rural, semi-urban and urban scenarios, respectively, before missing the deadline for a critical DAG. Furthermore, \SCHED is able to achieve an average of 2.4$\times$ better PE utilization over the baselines. This results in the SoC consuming an average of 3.5$\times$ lower energy when used with \SCHED in comparison to the prior schedulers.

\SCHED achieves 1.9--5.5$\times$ improvement in mission time and 1.8--4.0$\times$ improvement in energy over the state-of-the art schedulers for 3D Mapping, as shown in Figure~\ref{fig:real_world}(b). 
Additionally, at the maximum safe speed achieved by \SCHED, ADS and 2lvl-EDF are each able to complete only a maximum of 9\% of the mission before failing for the first critical DAG. Moreover, \SCHED achieves up to 1.7$\times$ better PE utilization. This improvement is lower than that for ADSuite, because Mapping has a high CPU utilization and low accelerator and GPU utilization, since many tasks execute only on the CPU.

The mission time for Package Delivery is shown in Figure~\ref{fig:real_world}(c). \SCHED achieves 1.9--4.7$\times$ improvement in mission time and 1.9--2.2$\times$ improvement in energy over the baseline schedulers. In terms of \% mission completion (not shown), ADS achieves a maximum of 38\% mission at the maximum safe speed of \SCHED for $Sys_B$. \SCHED is able to achieve up to 2.8$\times$ better average PE utilization for $Sys_B$.

We note that many of the tasks executed in 3D Mapping and Package Delivery only have CPU implementations (Table~\ref{tab:exec_times}). As developers implement heterogeneous algorithms for these tasks, we expect \SCHED to show greater benefits in terms of QoM metrics and PE utilization over the baseline schedulers. Moreover, \SCHED with \texttt{\MSdyn} performs significantly better for ADSuite in comparison to the baseline schedulers, as the deadlines for this application are more stringent ($\sim$400~ms) in comparison to MAVBench ($\sim$2000~ms).

\subsubsection{Idle Time Comparison}

\begin{figure}[t]
\centering
  \includegraphics[width=1\columnwidth]{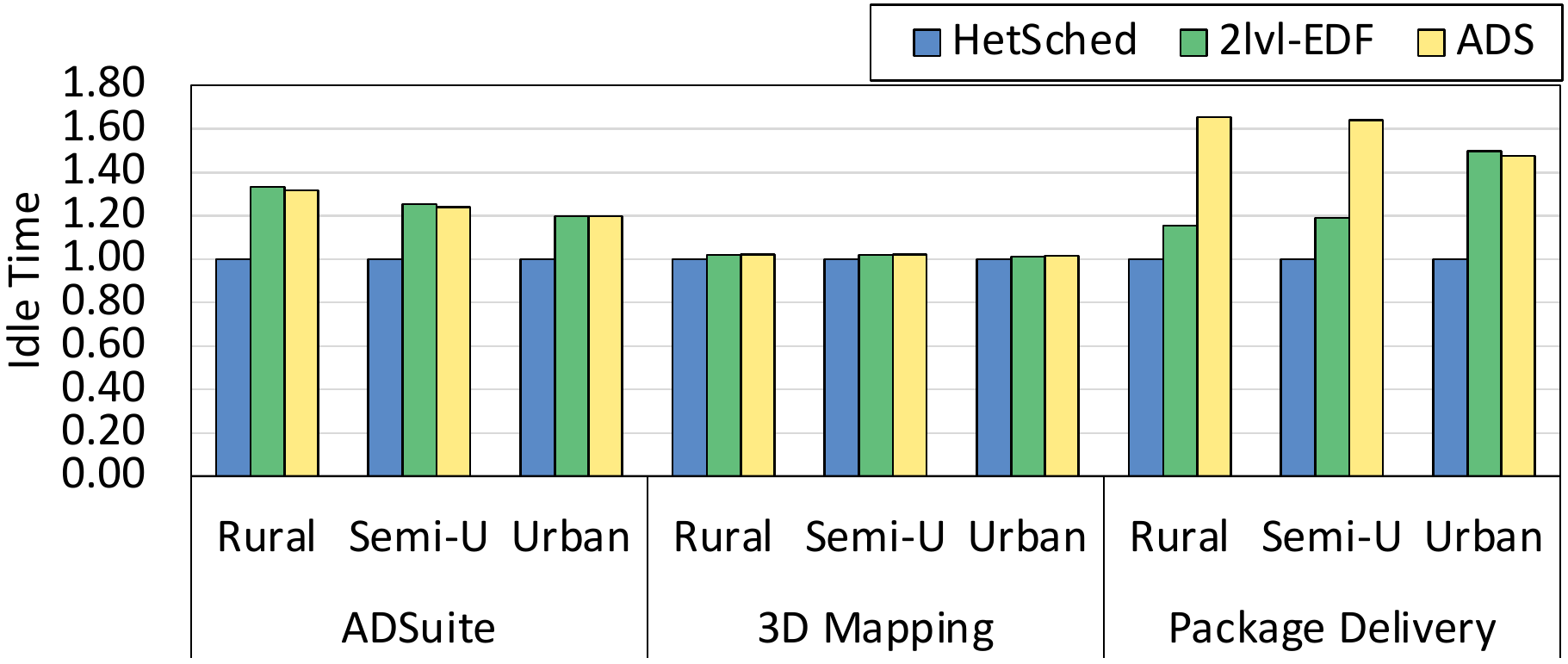}

  \caption{Idle time on $Sys_B$ using \SCHED, 2lvl-EDF and ADS when the system is operating at the speed at which all schedulers complete their mission safely.}
\label{fig:idle_time}
\end{figure}

To provide insight into the gains of \SCHED against ADS and 2lvl-EDF, we analyze the efficient use of resources by each scheduler when operated at a speed at which all schedulers can meet deadlines for all \texttt{crit=2} DAGs, i.e. at the speed that the worst scheduler-based system has to operate for safe mission completion. We evaluate the efficiency of resource management by comparing the idle time of all PEs for varying congestion scenarios for each scheduler. Note that we omit the analyses of the CPATH scheduler as the idle time for this scheduler was very small and did not allow for comparison against the real time schedulers.

For ADSuite, \SCHED achieves 33\% and 32\% higher idle time in comparison to 2lvl-EDF and ADS, respectively, for the rural scenario. As the congestion increases, the difference in the idle time between \SCHED and 2lvl-EDF and ADS reduces as shown in Figure~\ref{fig:idle_time}. For 3D mapping, the idle time difference is minimal as the tasks are only run on the CPU and GPU and do no have accelerated tasks implemented yet. For package delivery, as the congestion increases, the idle time decreases for 2lvl-EDF and remains almost same for ADS. 
\SCHED is, therefore, able to increase the speed of the AV and process more tasks per unit time. This increase in idle time can also be used to reduce power consumption of the system if there are external constraints on the speed at which the AV can operate.

\subsubsection{Scheduler Overhead}
We also evaluated the overhead of \SCHED, \ie the time spent for dependency tracking, meta information update, task prioritization and task assignment, running on the host Arm processor on the TX1. {We observed this overhead to be no more than 19\% and 6\% of the total mission time for ADSuite and MAVBench, respectively.}

\subsubsection{Energy and Available Slack}

\begin{figure}[t]
\centering

  \includegraphics[width=1\columnwidth]{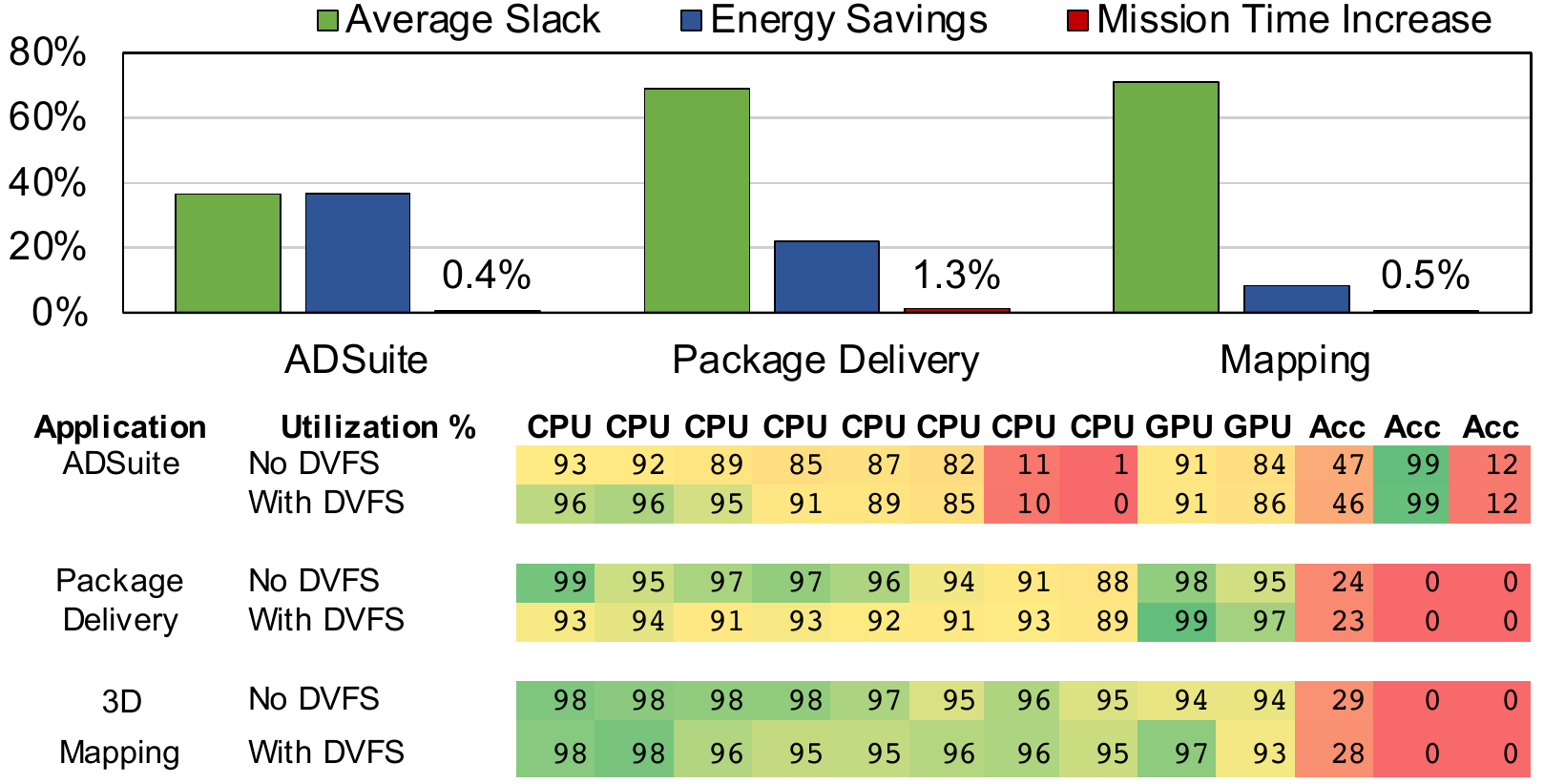}

  \caption{\textit{Top.} Slack, energy savings and mission overhead for \SCHED with DVFS enabled across the three applications. \textit{Bottom.} Per-PE utilization for each application. Results are shown for the rural congestion. The execution run corresponding to the $f_\text{slack}$ value with best energy savings is reported here. Note that the utilization values for two accelerators (DET and TRA) are zero because the latter two applications do not have any tasks that use these PEs.}
\label{fig:energy_slack}

\end{figure}

As discussed in Section~\ref{subsec:energy_power_model}, we adopt a scheme that dynamically recuperates a fraction ($f_\text{slack}$) of the slack savings enabled by \SCHED on a per-task basis (Section~\ref{subsec:energy_power_model}). Figure~\ref{fig:energy_slack} (top) shows that our DVFS policy allows \SCHED to achieve energy savings of 36\%, 22\% and 8\% for ADSuite, 3D Mapping and Package Delivery, respectively, while increasing the mission time by just 0.4--1.3\%. Note that this is at the maximum safe speed of the AV. At 85\% of the AV's maximum safe speed, we observe energy savings of up to 46.6\%, with average PE utilization improvement of up to 18.9\% (not shown).

We also show the per-PE utilization with and without DVFS in Figure~\ref{fig:energy_slack} (bottom), for each of the three applications. Much of the overall energy savings comes from DVFS on the CPU cores, since the CPU is typically the slowest and consumes the most power (Table~\ref{tab:exec_times}). DVFS improves the utilization of CPUs by 4\% on average for ADSuite, yielding the highest energy savings among the applications. DVFS may also reduce the utilization of a subset of the slower cores whenever there are changes to the schedule such that tasks that previously were executed on these PEs now migrate to a faster PE. We observe this for Package Delivery and Mapping, where migration of tasks from CPU to GPU contributes to reduced utilization for the slower CPUs, and yet lowers the overall energy consumption.

\subsubsection{\red{SoC Design Optimization}}
\begin{figure}[t]
\centering

  \includegraphics[width=1\columnwidth]{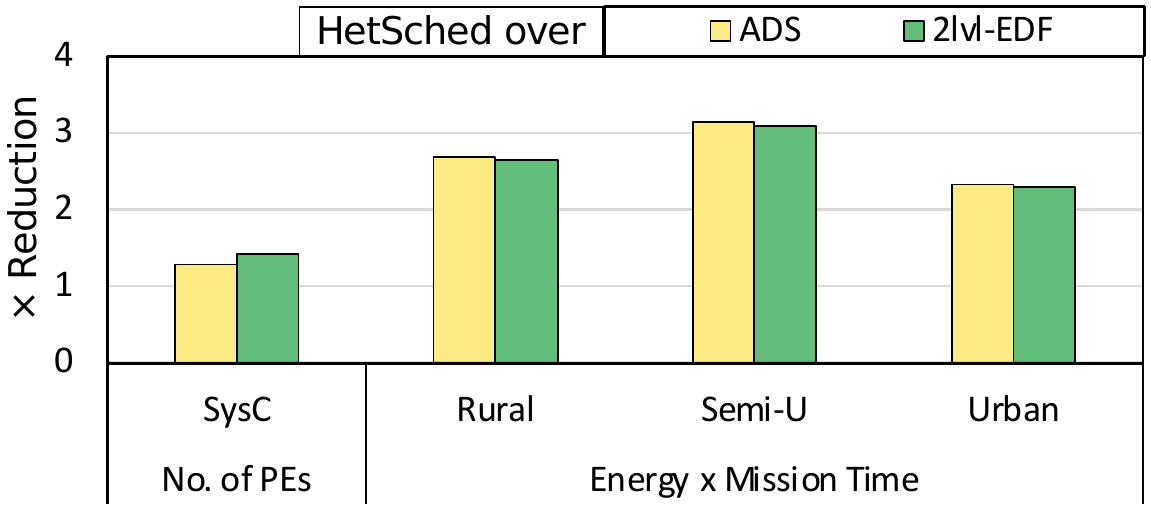}
  \vspace{-\baselineskip}

  \caption{\textit{Left:} Reduction in number PEs in $SysC$ when scheduler-in-the-loop design space exploration is performed. \textit{Right:} Improvement of \SCHED in product of energy and mission time for varying congestion levels for ADSuite over prior schedulers for respective $SysC$ configurations.}
\label{fig:optimizer}

\end{figure}

\red{As described in Section~\ref{sec:introduction}, having a scheduler-in-the-loop enables design space exploration to determine the best architectural configuration and level-of-heterogeneity for a given AV application. We used \SCHED to determine the best SoC configuration that optimizes upon mission time and energy consumption when executing AV applications in the urban scenario. We explore a set of design points, and pick the best SoC configuration as the one with the minimum energy--mission time product and smallest number of PEs (best PE utilization). We present only the results of ADSuite for brevity. The best SoC design configurations, called $SysC$, are $(16,4,2,4,1)$, $(16,8,2,4,1)$ and $(12,4,2,3,0)$ for ADS, 2lvl-EDF and \SCHED, respectively, where $(A,B,C,D,E)$ denotes $A$ detection accelerators, $B$ tracking accelerators, $C$ localization accelerators, $D$ GPUs and $E$ CPU cores. As shown in Figure~\ref{fig:optimizer}, for a reduction in number of PEs by 35\%, we are able to achieve an energy-mission time product reduction of 2.7$\times$ on average over ADS and 2lvl-EDF when operating in varying congestion levels.}

\section{Related Work}
\label{related_work}

AVs pose a challenge with the need to execute heterogeneous applications within stringent real-time and safety constraints. Prior work proposes the use of heterogeneous SoCs to help meet the performance constraints of individual tasks within the heterogeneous applications~\cite{lin2018architectural, teslafsd}. However, during runtime, multiple critical applications and tasks are required to be executed simultaneously within their deadlines~\cite{rt-survey}. 

A plethora of work exists on scheduling algorithms for heterogeneous systems. Much of the prominent schedulers focus on optimizing the makespan, \ie execution time of a single DAG~\cite{heft, cats, cpath}. Tong \etal use Q-learning along with heterogeneous earliest finish time (HEFT) algorithm from \cite{heft} to reduce the makespan of a DAG~\cite{ql_heft}. Shetti \etal propose HEFT-NC~\cite{heft-nc} to optimize ranking and task selection of HEFT~\cite{heft} by considering global and local processor information. However, as these schedulers are not optimized for the real-time requirements of the AVs and are not built for multiple DAG execution, they would need to be operated on AVs running at very low speeds to meet deadlines for all the critical DAGs of the mission.

To schedule for a multi-DAG scenario on heterogeneous systems, Xu \etal develop the reverse HEFT scheduling algorithm~\cite{rheft}. However, this algorithm is not feasible for dynamic systems as it requires \textit{a priori} knowledge of arrival times of all DAGs. Real-time schedulers like earliest deadline first (EDF) and deadline monotonic (DM), in contrast, cater to real-time systems where all tasks have a fixed priority and the criticality of tasks is not considered~\cite{rt-survey}. However, AVs are categorized as cyber-physical systems and require schedulers that can schedule for mixed criticalities and multiple DAGs~\cite{capota2019towards}. In this regard, Xie \etal~\cite{xie2017adaptive} propose two dynamic schedulers; fairness-based dynamic scheduler (FDS\_MIMF) and an adaptive dynamic scheduler (ADS\_MIMF), to optimize the makespan of the DAGs and to achieve low deadline miss ratio (DMR) by considering safety and criticality of the system  values for high-criticality DAGs, respectively. Wu and Ryu~\cite{wu2017bsf_edf}, present the best speed fit EDF scheduler that prioritizes tasks according to the earliest deadline and assigns the task to the best possible PE while considering the execution profile of the task, similar to the 2lvl-EDF implementation in this work. However, none of these work consider that meeting real-time deadlines does not translate to safe completion of mission at the least mission time, \ie while operating the AV at the maximum safe speed. Moreover, use of HEFT-like algorithms for task scheduling on a highly heterogeneous SoC leads to low utilization in slow PEs. \SCHED caters to both the requirements of an AV, \ie to meet real-time deadlines for critical DAGs and reduce overall mission time.

Recent work also propose the use of power management systems to help reduce the power and energy consumption of the real-time systemsusing machine learning techniques~\cite{HiLITE, mandal2020energy}. \SCHED can work in conjunction with these schedulers to further reduce energy along with the efficient utilization of resources to meet real-time deadlines and power constraints.

\section*{Acknowledgement}
This research was developed with funding from the Defense Advanced Research Projects Agency (DARPA). The views, opinions and/or other findings expressed are those of the authors and should not be interpreted as representing the official views or policies of the Department of Defense or the U.S. Government. 

\section{Conclusion}
\label{conclusion}

We presented a multi-level scheduler called \SCHED that exploits the highly heterogeneous nature of the underlying domain-specific systems-on-chip (DSSoC) in conjunction with the characteristics of an AV application. \SCHED's goal is to improve a global objective function, exemplified by a defined \textit{Quality-of-Mission} (QoM) metric, providing
a more \textit{holistic} scheduling approach that looks into the full hardware-software AV stack to improve the overall mission's quality rather than focusing solely on the real-time requirements of individual kernels or applications. Our evaluation shows that \SCHED on average improves the mission performance by 2.6--4.6$\times$  compared to state-of-the-art real-time heterogeneous schedulers. This is achieved with an average of {53.3\%} higher hardware utilization, while meeting {100\%} of critical deadlines on real-world AV applications. \SCHED can reduce the number of processing elements required in an SoC to safely complete AV missions by 35\%, while reducing the energy-mission time product by 2.7$\times$, when compared with prior schedulers for AV applications.

\bibliographystyle{IEEEtranS}
\bibliography{references}

\end{document}